\documentclass[fleqn,usenatbib]{mnras}

\usepackage{newtxtext,newtxmath}

\usepackage[T1]{fontenc}

\DeclareRobustCommand{\VAN}[3]{#2}
\let\VANthebibliography\thebibliography
\def\thebibliography{\DeclareRobustCommand{\VAN}[3]{##3}\VANthebibliography}

\usepackage{graphicx}	
\usepackage{amsmath}	
\usepackage{braket}

\title[On the time lags of V709 Cas]{On the frequency-dependent time lags of the intermediate polar V709 Cas}

\author[Z. A. Irving et al.]{
Zackery A. Irving,$^{1}$\thanks{E-mail: z.irving@soton.ac.uk}
Noel Castro Segura,$^{2}$
Diego Altamirano,$^{1}$
Martina Veresvarska,$^{3,4}$
\newauthor
Mariano M\'endez,$^{5}$
Federico Garc\'ia,$^{6}$
Domitilla de Martino,$^{7}$
Simone Scaringi,$^{8,7}$
Federico Vincentelli,$^{9}$
\newauthor
Angel Castro,$^{10}$
Federico Soto,$^{10}$
Raul Michel$^{10}$
\\
$^{1}$School of Physics and Astronomy, University of Southampton, University Road, Southampton SO17 1BJ, UK \\
$^{2}$Department of Physics, University of Warwick, Gibbet Hill Road, Coventry CV4 7AL, UK\\
$^{3}$Institute of Space Sciences (ICE, CSIC), Campus UAB, Carrer de Can Magrans s/n, E-08193 Barcelona, Spain\\
$^{4}$Institut d’Estudis Espacials de Catalunya (IEEC), 08860 Castelldefels (Barcelona), Spain\\
$^{5}$Kapteyn Astronomical Institute, University of Groningen, PO Box 800, CH-9700 AV Groningen, the Netherlands\\
$^{6}$Instituto Argentino de Radioastrono\'ia (CCT La Plata, CONICET; CICPBA; UNLP), C.C.5, (1894) Villa Elisa, Buenos Aires, Argentina\\
$^{7}$INAF -- Osservatorio Astronomico di Capodimonte, Salita Moiariello 16, I-80131 Napoli, Italy\\
$^{8}$Centre for Extragalactic Astronomy, Department of Physics, Durham University, South Road, Durham DH1 3LE, UK\\
$^{9}$Fluid and Complex Systems Centre, Coventry University, CV1 5FB, UK\\
$^{10}$Instituto de Astronom\'ia, Universidad Nacional Autónoma de M\'exico, AP 106,  Ensenada 22800, BC, M\'exico
}

\date{Accepted XXX. Received YYY; in original form ZZZ}

\pubyear{\the\year{}}

\begin{document}
\label{firstpage}
\pagerange{\pageref{firstpage}--\pageref{lastpage}}
\maketitle

\begin{abstract}

Cataclysmic variables (CVs) are binary systems in which a white dwarf (WD) primary accretes material from a late-type secondary, typically via a disc. To date, only six CVs, all of which are non-magnetic, have been found to exhibit lags (delayed variability in one band with respect to another). In all six cases, these lags are ``red'' (i.e., red lags blue) with a time lag of order seconds. While there is no generally accepted mechanism for producing red lags in CVs, current theories suggest reprocessing in the disc, either on the thermal or recombination time-scales, or inside-out shocks propagating through the disc. Using $g$-, $r$-, and $i$-band data from the OPtical TIming CAMera (OPTICAM), we report the discovery of frequency-dependent red lags from V709 Cas: a magnetic CV of the intermediate polar (IP) subclass. These red lags reach a maximum of $\sim$4--6~s on time-scales of 7.4--13.0~min, similar to previously-reported lags in non-magnetic CVs. The detection of lags in an IP is significant as the WD's magnetic field truncates the accretion disc, limiting disc-based lag mechanisms to large radii; as a result, characteristic accretion disc time-scales cannot explain the observed lags. However, recombination can occur on second time-scales, even in systems with truncated discs. For example, we show that recombination in the disc's bright spot, where overflowing material from the secondary meets the outer-edge of the disc, plausibly explains V709 Cas's optical spectrum, the prominence of optical spin-orbit beat pulsations, and the red lags found in this work.

\end{abstract}




\section{Introduction}\label{sec: intro}

Cataclysmic variables (CV) are binary systems consisting of a white dwarf (WD) primary and a late-type, Roche-lobe-filling secondary \citep[e.g.,][]{Warner1995}. Roche lobe overflow leads to material from the secondary being accreted onto the primary \citep[e.g.,][]{Kopal1959}. To conserve angular momentum, overflowing material from the secondary typically forms an accretion disc around the WD. If the WD is moderately magnetic ($B \lesssim 10$~MG), the accretion disc may be truncated at the magnetospheric radius; such systems are referred to as intermediate polars (IPs; see, e.g., \citealt{Patterson1994} for a review). Since the disc of an IP is truncated, material cannot be accreted directly onto the WD surface. Instead, material is magnetically confined onto the magnetic poles of the WD, allowing pulsations at the WD spin period to be observed \citep[e.g.,][]{Walker1956, Walker1958, Walker1961, Katz1975}. If the WD is strongly magnetised ($B \gtrsim 10$~MG), the magnetospheric radius may be so large as to prevent the formation of an accretion disc \citep[e.g.,][]{King1979}. Another characteristic feature of strongly magnetic CVs is synchronism: the binary orbital period is equal to the WD spin period \citep[e.g.,][]{Chanmugam1977} - though a small number of asynchronous systems have been observed \citep[e.g.,][]{Patterson1995, Rea2017, Tovmassian2017}. CVs with strongly magnetic WDs are referred to as polars (see \citealt{Cropper1990} for a review).

IPs are a particularly interesting subclass of CVs due to their high magnetic moment, and thus large magnetospheres, while still possessing an accretion disc (e.g., \citealt{Warner1995}). Moreover, IPs have been found to exhibit a number of phenomenological similarities to low-mass X-ray binaries \citep[LMXBs; e.g.,][]{Warner2004}. A commonly observed phenomenon of X-ray binaries (XRBs) is the presence of frequency-dependent lags: variability at a particular frequency seen in one energy band is also seen in another energy band after some delay (e.g., \citealt{Miyamoto1988, Nowak1999}). Below $\sim 10$~Hz, XRBs may exhibit so-called ``hard'' lags in which hard (i.e., high energy) photons lag soft (i.e., low energy) photons, typically by tens or hundreds of milliseconds \citep[e.g.,][]{Kotov2001, Cassatella2012}. The physical mechanism producing these lags is believed to be propagating fluctuations in the accretion disc \citep{Lyubarskii1997, Arevalo2006, Ingram2011}, or inverse Compton scattering in the X-ray corona \citep[e.g.,][]{Karpouzas2020, Karpouzas2021} or jet \citep[e.g.,][]{Reig2003, Giannios2004, Giannios2005}. Within the propagating fluctuations model, perturbations originate in the disc at large radii and propagate inwards: at smaller radii the disc is hotter, thereby producing harder photons which lag the softer photons produced at larger radii. Conversely, (soft) disc photons may be observed directly, followed by hard photons which have been Compton upscattered in the corona/jet. In addition, XRBs may also exhibit ``soft'' lags, in which the soft photons lag hard photons. Soft lags in XRBs are typically associated with high-frequency QPOs \citep[e.g.,][]{Vaughan1998, Kaaret1999, Reig2000, Karpouzas2020}, and can be explained by reverberation, similarly to AGN \citep[e.g.,][]{Fabian2009, Zoghbi2010}, or feedback from the corona on to the disc \citep[e.g.,][]{Karpouzas2020, Karpouzas2021}.

For XRBs, the cross-spectrum between two (X-ray) light curves has become the standard method for measuring lags (see, e.g., \citealt{Uttley2014} for a review). Unlike the cross-correlation function (CCF), the cross spectrum can be used to measure lags at specific Fourier frequencies, isolating properties of the variability over certain time-scales. As such, the cross spectrum can be used to identify lags that might otherwise be missed by the CCF \citep[e.g.,][]{Papadakis2001}. Additionally, the cross spectrum can be used to compute the coherence between two light curves, which may be interesting in and of itself \citep[e.g.,][]{Vaughan1997}. Recently, \cite{Mendez2024} showed that ``hidden'' variability components can be found by jointly-modelling the cross and power spectra between two light curves; in some cases, these hidden components are required to reproduce the coherence and lag spectrum. 

In X-rays, searches for lags can cover several orders of magnitude in frequency due to the high time resolution of X-ray telescopes. Moreover, X-ray detectors typically measure photon energies in addition to their times of arrival, enabling construction of simultaneous light curves in multiple energy bands from a single instrument. In the case of CVs, however, most of the emission is in the optical regime \citep[e.g.,][]{Warner1995}, and the time-scales of variability are typically much longer \citep[e.g.,][]{Belloni2002, Revnivtsev2010}. As such, searching for frequency-dependent lags in CVs is more challenging for a number of reasons. Firstly, optical instruments lack the time resolution of their X-ray counterparts. Secondly, obtaining simultaneous optical light curves in multiple bands requires simultaneous observations - either using multiple independent instruments or multi-camera instruments like ULTRACAM \citep{Dhillon2007}. Finally, most optical telescopes are ground-based, limiting the lengths of the observations. As such, the study of frequency-dependent time lags in CVs is much more limited than in XRBs, and lags have only been found in a handful of sources.

Using short-cadence simultaneous three-colour observations from ULTRACAM \citep{Dhillon2007}, \cite{Scaringi2013} were the first to identify frequency-dependent lags in CVs. Specifically, \cite{Scaringi2013} found low-frequency (i.e., $< 10^{-2}$~Hz) ``red'' lags, in which variability at redder wavelengths is delayed with respect to variability at bluer wavelengths, with time lags of $\sim 3$~s and $\sim 10$~s for MV Lyr and LU Cam, respectively. To explain these lags, \cite{Scaringi2013} suggested reprocessing in the disc on the thermal time-scale. Alternatively, \cite{Scaringi2013} suggested that the observed lags could be result of reverse (inside-out) shocks, effectively the reverse of the hard-lag-producing propagating fluctuations model for XRBs \citep[e.g.,][]{Lyubarskii1997, Kotov2001, Arevalo2006}.

Red frequency-dependent lags of up to $\sim 5$~s, occurring on a time-scale of $\sim 250$~s, were also seen in the ULTRACAM data of SS Cyg \citep{Aranzana2018}. \cite{Aranzana2018} considered the thermal time-scale reprocessing mechanism proposed by \cite{Scaringi2013} and found that, while plausible, this mechanism required high irradiation and high viscosity. As an alternative, \cite{Aranzana2018} suggested that the lags could be a result of reprocessing on the recombination time-scale in the upper-layers of the disc. Reprocessing on the recombination time-scale has also been invoked to explain the lag in the optical emission of XRBs with respect to their X-ray emission \citep[e.g.,][]{OBrien2002}. Frequency-independent red lags have also been detected in V603 Aql ($17 \pm 7$~s), TT Ari ($2.3 \pm 0.5$~s), and RS Oph ($10 \pm 1$~s), using data from a number of different instruments \citep{Bruch2015}. In contrast to XRBs, there have so far been no instances of CVs exhibiting statistically significant blue lags.

V709 Cassiopeiae (hereafter V709 Cas) is an IP that was discovered in the \textit{ROSAT} all-sky survey \citep[][]{Haberl1995, Motch1996}. Despite clear and consistent WD spin pulsations being seen in X-rays \citep[e.g.,][]{Norton1999, DDM2001, Mukai2015}, V709 Cas's WD spin pulsations have historically proven difficult to observe in the optical. A few years after V709 Cas's discovery, \cite{Kozhevnikov2001} found optical spin and spin-orbit beat pulsations, while later optical studies were unable to reliably reproduce these results \citep[e.g.,][]{Tamburini2009, Hric2014}. In the red optical--near-infrared, \cite{Rao2026} were able to determine precise WD spin, spin-orbit beat, and binary orbital periods of $312.7478(2)$~s\footnote{The parentheses denote the uncertainty on the final digit.}, $317.9267(2)$~s, and $5.3329(2)$~hr, respectively, using long baseline data from the \textit{Transiting Exoplanet Survey Satellite (TESS)}. \cite{Rao2026} also found that the relative amplitudes of the spin and spin-orbit beat pulsations varied over time, leading them to suggest that V709 Cas exhibits a changing accretion geometry: when the spin is dominant, the WD is accreting via a disc; when the beat is dominant, the WD is accreting via a stream. Using simultaneous three-colour observations in the SDSS $g$-, $r$-, $i$-, and $z$-bands, however, \cite{Irving2026} found that the relative amplitudes of the spin and beat pulsations varied with wavelength: in the optical $g$- and $r$-bands, the beat is prominent while the spin is not; in the near-infrared $i$- and $z$-bands, the spin is prominent while the beat is not. Instead of a varying accretion geometry, \cite{Irving2026} therefore suggested that V709 Cas may emit near-infrared cyclotron radiation from its accretion columns.

In this work, we report the discovery of red lags in the optical emission of V709 Cas using data collected by the OPtical TIming CAMera (OPTICAM; \citealt{Castro2019, Castro2024}). OPTICAM is a high-cadence triple-camera system mounted on the 2.1~m telescope at the Observatorio Astron\'omico Nacional in San Pedro M\'artir (OAN-SPM), M\'exico. To our knowledge, this is the first time lags have been reported for an IP; V709 Cas is particularly interesting as a candidate for cyclotron emission in the near-infrared. The structure of this work is as follows: Section \ref{sec: data} details our observations; Section \ref{sec: methods} describes the analyses performed; the results of these analyses are then presented in Section \ref{sec: results}; in Section \ref{sec: discussion}, we attempt to interpret our results; finally, our closing remarks are given in Section \ref{sec: conclusions}.

\section{Data}\label{sec: data}

\subsection{OPTICAM}\label{sec: opticam data}

\begin{figure*}
    \centering
    \includegraphics[width=\textwidth]{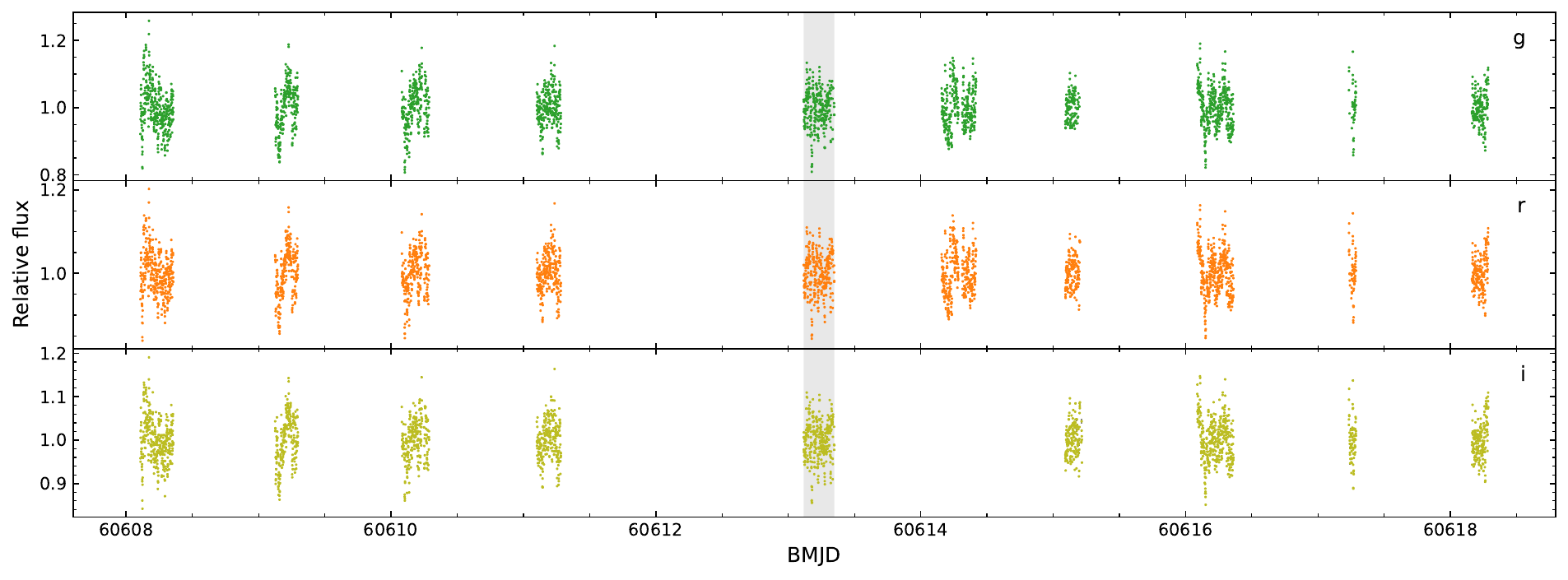}
    \includegraphics[width=\textwidth]{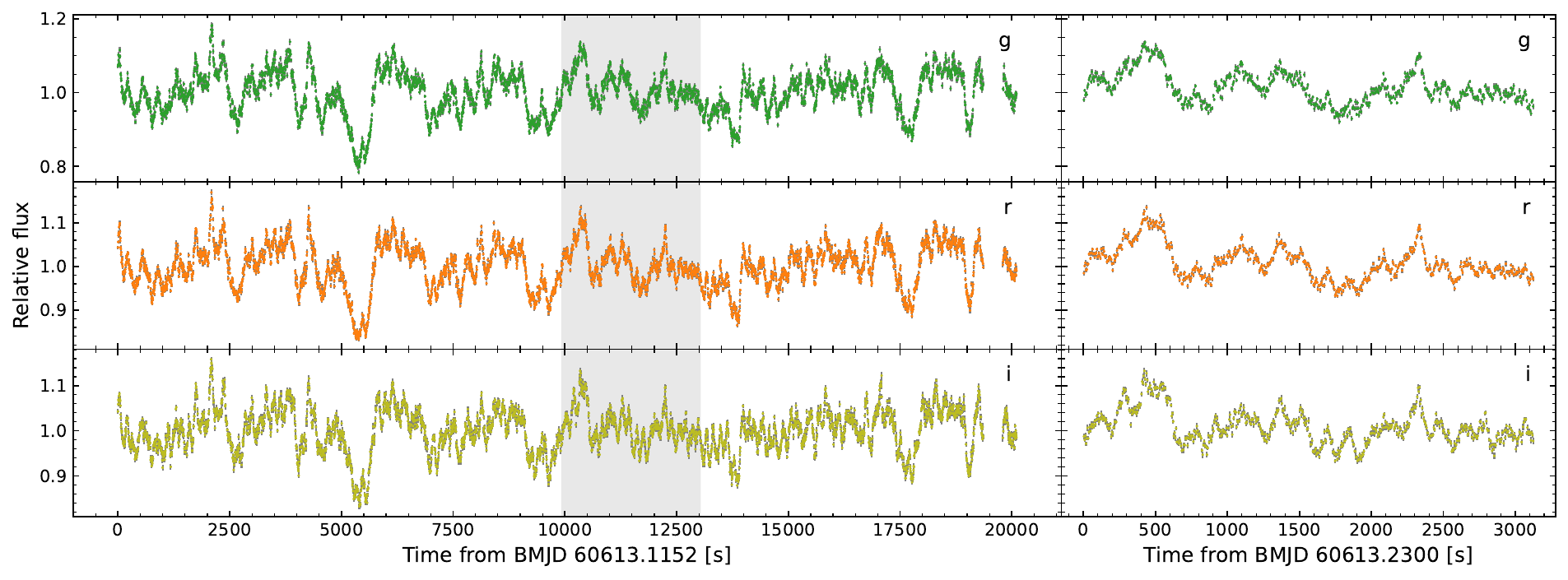}
    \caption{Top panel: all OPTICAM light curves of V709 Cas binned to a time resolution of 60~s. Bottom left panel: Epoch 5 (corresponding to the shaded grey region in the top panel) binned to a time resolution of 5~s. Bottom right panel: zoomed 3125~s inset of Epoch 5 (corresponding to the shaded grey region in the bottom left panel).}
    \label{fig: light curves}
\end{figure*}

OPTICAM observed V709 Cas in the $g$-, $r$-, and $i$-/$z$-bands over 10 nights in late October, 2024, through early November, 2024. These observations are detailed in Table \ref{tab: observations}, and the resulting light curves are presented in Figure \ref{fig: light curves}. The data analysed in this work is a subset of the data analysed in \cite{Irving2026}; for details on data reduction, we therefore refer to Section 2.1 of \cite{Irving2026}.

\begin{table}
    \centering
    \caption{OAN-SPM 2.1~m observations of V709 Cas.}
    \begin{tabular}{c|c|c|c|c}
        \hline
        OPTICAM & Date & Filters & Cadence & Exposure \\
        epoch & (YYYY-MM-DD) &  & [s] & [s] \\
        \hline
        1 & 2024-10-25 & $g$, $r$, $i$ & 3 & 21330 \\
        2 & 2024-10-26 & $g$, $r$, $i$ & 3 & 14718 \\
        3 & 2024-10-27 & $g$, $r$, $i$ & 3 & 16104 \\
        4 & 2024-10-28 & $g$, $r$, $i$ & 3 & 15507 \\
        5 & 2024-10-30 & $g$, $r$, $i$ & 3 & 19278 \\
        6 & 2024-10-31 & $g$, $r$, $z^*$ & 3 & 19827 \\
        7 & 2024-11-01 & $g$, $r$, $i$ & 3 & 9885 \\
        8 & 2024-11-02 & $g$, $r$, $i$ & 3 & 23136 \\
        9 & 2024-11-03 & $g$, $r$, $i$ & 3 & 4692 \\
        10 & 2024-11-04 & $g$, $r$, $i$ & 3 & 10503 \\
        \hline
        B \& C & Date & \multicolumn{2}{c}{Wavelength range} & Exposure \\
         & (YYYY-MM-DD) & \multicolumn{2}{c}{[\r{A}]} & [s] \\
        \hline
         & 2026-06-30 & \multicolumn{2}{c}{4200--7460} & 300 \\
        \hline
    \end{tabular}
    {\\NOTES: $^*$$z$-band data are not included in subsequent analyses.}
    \label{tab: observations}
\end{table}

OPTICAM's hardware generated time-stamps are accurate to within 25~ns \citep{Castro2019}, though clock synchronisation introduces a delay of up to $\sim20$--40~ns between cameras. We note, however, that a full characterisation of OPTICAM's timing accuracy, including instrumental latency and rolling shutter time skew, has not yet been performed. As such, OPTICAM's timing accuracy is currently limited to 1~ms by its software timestamps.

\subsection{Boller \& Chivens spectrograph}\label{sec: spectrum data}

To complement our OPTICAM observations, we also observed V709 Cas using the Boller \& Chivens spectrograph on the 2.1~m telescope at the OAN-SPM on June 30, 2026. We used a 400~g/mm grating with a blaze angle of $6^\circ 30$', set to a grating angle of $7^\circ 25$', to obtain low-resolution ($7.65 \pm 0.72$~\r{A} FWHM) spectroscopy in the 4200--7460~\r{A} range using a 200-micron slit. The exposure time was 300~s, roughly matching V709 Cas's 312.74~s WD spin period \citep[e.g.,][]{Rao2026}, and we used a 2$\times$2 binning. This observation is also listed in Table \ref{tab: observations}. The spectrum was wavelength-calibrated using a CuAr lamp.

Since our optical spectrum was taken more than a year after V709 Cas was observed by OPTICAM, it is instructive to review V709 Cas's optical flux during these observations. During the OPTICAM observing run, V709 Cas had an average All-Sky Automated Survey for SuperNovae (ASAS-SN) $g$-band magnitude of 14.8 \citep[][]{Shappee2014, Kochanek2017}. Similarly, the day before our optical spectrum was taken,\footnote{ASAS-SN did not observe V709 Cas on June 30, 2026.} V709 Cas had an average ASAS-SN $g$-band magnitude of 14.5. Despite the large time gap, V709 Cas therefore had a similar optical flux during both sets of observations.

To flux-calibrate our optical spectrum, we first applied bias and flat-field corrections. For flat-fielding, we used an internal halogen lamp to correct for pixel-to-pixel sensitivity variations, and twilight flats of the sky to correct for large-scale variations along the slit. The sky background was subtracted using sampling regions free of stellar emission, but we did not correct for telluric absorption. Finally, the spectrum was flux-calibrated using the spectrophotometric standard star Feige 110 \citep{Massey1988}.

We also corrected our spectrum for interstellar dust extinction (i.e., ``reddening'') following \cite{ODonnell1994}. We determined the extinction, $A_V$, using the X-ray spectral fits from \cite{Irving2026}, resulting in $A_V = 0.44$ \citep{Predehl1995}. For the ratio of total to selective extinction, $R_V \equiv A_V / E(B-V)$, where $E(B-V)$ is the difference in extinction between the $B$ and $V$ bands, we adopted $R_V = 3.1$ following \cite{Bastiaansen1992}.

\section{Methods}\label{sec: methods}

\subsection{Fourier}

\subsubsection{Power spectrum}

The power spectrum, or periodogram, of a time series, $x(t)$, is defined as:
\begin{equation}
    P_x(\nu) = |X(\nu)|^2,
    \label{eq: power spectrum}
\end{equation}
where $\nu$ represents frequency and $X(\nu)$ is the Fourier transform of $x(t)$. While $X(\nu)$ will be complex, it is clear that $P(\nu)$ is necessarily real-valued for all values of $\nu$. 

Since the estimate of any single power spectrum is noisy, it is common to average power spectra computed from non-overlapping time series segments \citep[e.g.,][]{vanderKlis1989}. Averaging power spectra in this way increases the signal-to-noise ratio at each frequency and allows uncertainties to be assigned to each power. The averaged power spectrum can be defined as:
\begin{equation}
    \braket{P_x(\nu)} = \braket{|X(\nu)|^2},
\end{equation}
where the angled brackets represent averages over independent power spectra. The uncertainties on the averaged powers are then given by:
\begin{equation}
    \delta \braket{P_x(\nu)} = \frac{\braket{P_x(\nu)}}{\sqrt{M}},
\end{equation}
where $M$ represents the number of power spectra that have been averaged. To ensure that the errors on the powers are meaningful, it is necessary to average a large number of power spectra.

To further improve the S/N, the averaged power spectrum can be binned in frequency \citep[e.g.,][]{vanderKlis1989}:
\begin{equation}
    \braket{\braket{P_x(\nu)}} = \braket{\braket{|X(\nu)|^2}},
\end{equation}
where we now have two sets of angled brackets to denote two averages: first, $M$ independent spectra are averaged; second $W$ adjacent frequency bins are averaged. It is common to use increasing bin sizes with increasing frequency - so-called logarithmic binning - and so $W$ is generally a function of the binned frequency, $\nu_w$. The errors on the binned powers are then given by:
\begin{equation}
    \delta \braket{\braket{P_x(\nu)}} = \frac{\braket{\braket{P_x(\nu)}}}{\sqrt{M W}}.
\end{equation}

\subsubsection{Cross spectrum}\label{sec: cross spectrum}

The cross spectrum between two time series, $x(t)$ and $y(t)$, is defined as:
\begin{equation}
    C(\nu) = X^*(\nu) Y(\nu),
    \label{eq: cross spectrum}
\end{equation}
where $X^*(\nu)$ is the complex conjugate of the Fourier transform of $x(t)$, and $Y(\nu)$ is the Fourier transform of $y(t)$ \citep[e.g.,][]{Vaughan1997}. Given this definition, it is clear that, unlike the power spectrum, the cross spectrum will be a complex quantity. As with the power spectrum, it is common to average and logarithmically bin the cross spectrum to improve its S/N:
\begin{equation}
    \braket{\braket{C(\nu)}} = \braket{\braket{X^*(\nu) Y(\nu)}}.
\end{equation}
For simplicity, we will only use one set of angled brackets for the remainder of this section.

From the averaged cross spectrum, the \textit{raw coherence} can be computed \citep{Bendat1986}:
\begin{equation}
    \gamma_\text{raw}^2 (\nu) = \frac{|\braket{C(\nu)}|^2}{\braket{|X(\nu)|^2} \braket{|Y(\nu)|^2}},
    \label{eq: raw coherence}
\end{equation}
with associated errors:
\begin{equation}
    \delta \gamma_\text{raw}^2 (\nu) = \sqrt{\frac{2}{MW}} \frac{\gamma_\text{raw}^2(\nu) (1 - \gamma_\text{raw}^2(\nu))}{\gamma_\text{raw}(\nu)}.
\end{equation}
Additionally, the phase lags can be computed directly from the averaged cross spectrum:
\begin{equation}
    \phi(\nu) = \arctan \Big( \frac{\text{Im}[\braket{C(\nu)}]}{\text{Re}[\braket{C(\nu)}]} \Big),
\end{equation}
with associated errors:
\begin{equation}
    \delta \phi (\nu) = \sqrt{\frac{1 - \gamma_\text{raw}^2 (\nu)}{2 \gamma_\text{raw}^2 (\nu) MW}}.
\end{equation}
The phase lags and their errors can be converted to units of time by dividing by angular frequency, $\omega \equiv 2 \pi \nu$:
\begin{equation}
    \tau (\nu) = \frac{\phi(\nu)}{\omega},
\end{equation}
where $\tau (\nu)$ represents the time lags. Similarly, the errors on the time lags are given by: 
\begin{equation}
    \delta \tau (\nu) = \frac{\delta \phi(\nu)}{\omega}.
\end{equation}

To compute the \textit{intrinsic coherence}, equation \ref{eq: raw coherence} must be corrected for both Poisson noise, and for the fact that a finite number of spectra have been averaged, both of which lead to a bias in the numerator \citep{Vaughan1997}:
\begin{equation}
    \gamma_\text{i} (\nu)^2 = \frac{|\braket{C(\nu)}|^2 - n(\nu)^2}{\braket{|X'(\nu)|^2} \braket{|Y'(\nu)|^2}},
    \label{eq: intrinsic coherence}
\end{equation}
where $n(\nu_k)^2$ is the bias term, $\braket{|X'(\nu)|^2} \equiv \braket{|X(\nu)|^2} - |N_x|^2$, where $|N_x|^2$ is the Poisson noise level of $\braket{|X(\nu)|^2}$, and $\braket{|Y'(\nu)|^2} \equiv \braket{|Y(\nu)|^2} - |N_y|^2$, where $|N_y|^2$ is the Poisson noise level of $\braket{|Y(\nu)|^2}$. The bias term is given by:
\begin{equation}
    n(\nu)^2 = \frac{\braket{|X'(\nu)|^2} |N_y|^2 + \braket{|Y'(\nu)|^2} |N_x|^2 + |N_x|^2 |N_y|^2}{MW}.
    \label{eq: bias term}
\end{equation}
The corresponding error on the intrinsic coherence is given by:
\begin{multline}
    \delta \gamma_i (\nu)^2 = \frac{\gamma_i (\nu)^2}{\sqrt{MW}} \times \\ \Big( \frac{2 n(\nu)^4 M W}{(|\braket{C(\nu)}|^2 - n(\nu)^2)^2} + \frac{|P_x|^4}{\braket{|X'(\nu)|^4}} + \\ \frac{|P_y|^4}{\braket{|Y'(\nu)|^4}} + \frac{2 (1 - \gamma_i(\nu)^2)^2}{\gamma_i(\nu)^3} \Big)^{1/2}.
\end{multline}

\subsubsection{Normalisation}\label{sec: normalisation}

When computing cross and power spectra, we use the fractional rms normalisation:
\begin{equation}
    P_\text{rms} (\nu) = \frac{2 \delta t}{\mu^2 N} |X(\nu)|^2,
    \label{eq: rms normalisation}
\end{equation}
where $\delta t$ represents the time resolution of the light curve, $\mu$ represents the mean flux, and $N$ represents the number of points in the light curve. This normalisation is defined such that the integrated power spectrum yields the square of the fractional rms of the light curve \citep[][]{Belloni1990}.

\subsubsection{Segment size and binning}\label{sec: segment size and binning}

We found that the timestamps of our observations exhibit a small amount of jitter, $\mathcal{O}(0.01)$~s, in the time between exposures. To avoid spectral leakage in our cross and power spectra, we therefore rebinned our OPTICAM light curves to a time resolution of 5~s. We then computed cross and power spectra using 3125~s segments, excluding incomplete segments and data gaps. A representative 3125~s light curve segment for each band is shown in the bottom right panel of Figure \ref{fig: light curves}. We also binned all of our spectra logarithmically using a scale factor of 1.02, such that each frequency bin was 2 per cent larger than the previous one.

\subsection{Multi-spectrum fitting}

To further test the similarities between CVs and XRBs, we apply the multi-spectrum modelling framework presented in \cite{Mendez2024} to a CV for the first time. Briefly, this method involves jointly-fitting a superposition of Lorentzians to the power spectra of two light curves in addition to the real and imaginary parts of their cross spectrum. To do this, we implement their constant phase lags model. From the joint cross and power spectral fit, the intrinsic coherence and lag spectrum can then be derived; while the intrinsic coherence may be an interesting quantity \citep{Vaughan1997}, we are primarily interested in testing whether this model can recover the lag spectra of CVs.

The multi-spectrum modelling framework presented in \cite{Mendez2024} assumes: i) the power spectrum of each light curve can be reproduced using a linear combination of Lorentzians\footnote{\cite{Mendez2024} note that this framework is not limited to Lorentzians, but they are conventionally used.}; ii) each Lorentzian has the same centroid frequency and full width at half maximum (FWHM) in all bands; iii) each Lorentzian is perfectly coherent in any two bands; iv) any two Lorentzians (which overlap in frequency) are totally incoherent with each other. For a discussion of these assumptions, we refer to \cite{Mendez2024}.

Since the cross spectrum is a complex quantity, each ordinate represents a vector in the complex plane. When fitting the cross spectrum, it is convenient to apply a global rotation, $\theta$, to these vectors; instead of fitting the real and imaginary parts of the cross spectrum, one then fits $\text{Re}(\braket{C(\nu)} \sin(\theta)) + \text{Im}(\braket{C(\nu)} \cos(\theta))$ and $\text{Re}(\braket{C(\nu)} \cos(\theta)) - \text{Im}(\braket{C(\nu)} \sin(\theta))$. By setting $\theta = 45$ degrees, all four spectra will have similar amplitudes, making fitting more practical. Moreover, since the angle of rotation is known, it is straightforward to recover the ``un-rotated'' real and imaginary parts of the cross spectrum (see \citealt{Mendez2024} for further details).

\subsubsection{The Lorentzian}

We define the Lorentzian as:
\begin{equation}
    L(\nu, A, \nu_0, \Delta) = A^2 \frac{\Delta}{\pi - 2 \arctan(-2 \nu_0 / \Delta)} \frac{1}{(\Delta / 2)^2 + (\nu - \nu_0)^2}
    \label{eq: Lorentzian}
\end{equation}
where $A$ represents the square-root of the area of the Lorentzian integrated from zero to infinity, $\nu_0$ is the centroid frequency, and $\Delta$ is the full width at half maximum (FWHM). Since we use the fractional rms normalisation (Section \ref{sec: normalisation}), $A$ is a direct measure of the fractional rms of the variability. For fitting, it can be convenient to replace $\Delta$ with $Q \equiv \nu_0 / \Delta$, which provides a more intuitive characterisation of a Lorentzian's width. Additionally, fixing $\nu_0$ to zero may provide a better fit to broad-band variability \citep[e.g.,][]{Belloni2002}; in this case, the Lorentzian is referred to as a zero-centred Lorentzian. When fitting a Lorentzian, we consider it to be significant if $A$ is more than $3 \sigma$ above zero.

\subsubsection{Optical caveats}\label{sec: optical caveats}

For the optical data analysed herein, the Nyquist frequency is 0.1~Hz. Unlike the X-ray data analysed in \cite{Mendez2024}, attenuation may therefore be significant \citep[e.g.,][]{vanderKlis1989}. Moreover, the power spectra of these light curves do not contain a high-frequency tail from which the ``Poisson'' noise level\footnote{We note that the noise in optical data is not strictly Poissonian, but we refer to it as such for consistency.} can be measured reliably. To estimate the Poisson noise level, we therefore include constant terms in our power spectrum models. We note, however, that without a high-frequency Poisson noise tail, our Poisson noise estimates may be skewed by attenuation. We recall that the intrinsic coherence and its errors are sensitive to the Poisson noise level, and so misestimation may skew these quantities. Poisson noise does not affect the cross spectrum unless there is ``cross-talk'' between the two light curves (e.g., due to overlapping energy ranges or correlated systematics). Since OPTICAM uses independent detectors for each filter, and there is minimal overlap between filters, we do not include a constant term in our cross spectrum models.

\section{Results}\label{sec: results}

\subsection{Power spectra}\label{sec: power spectra results}

\begin{figure}
    \centering
    \includegraphics[width=\columnwidth]{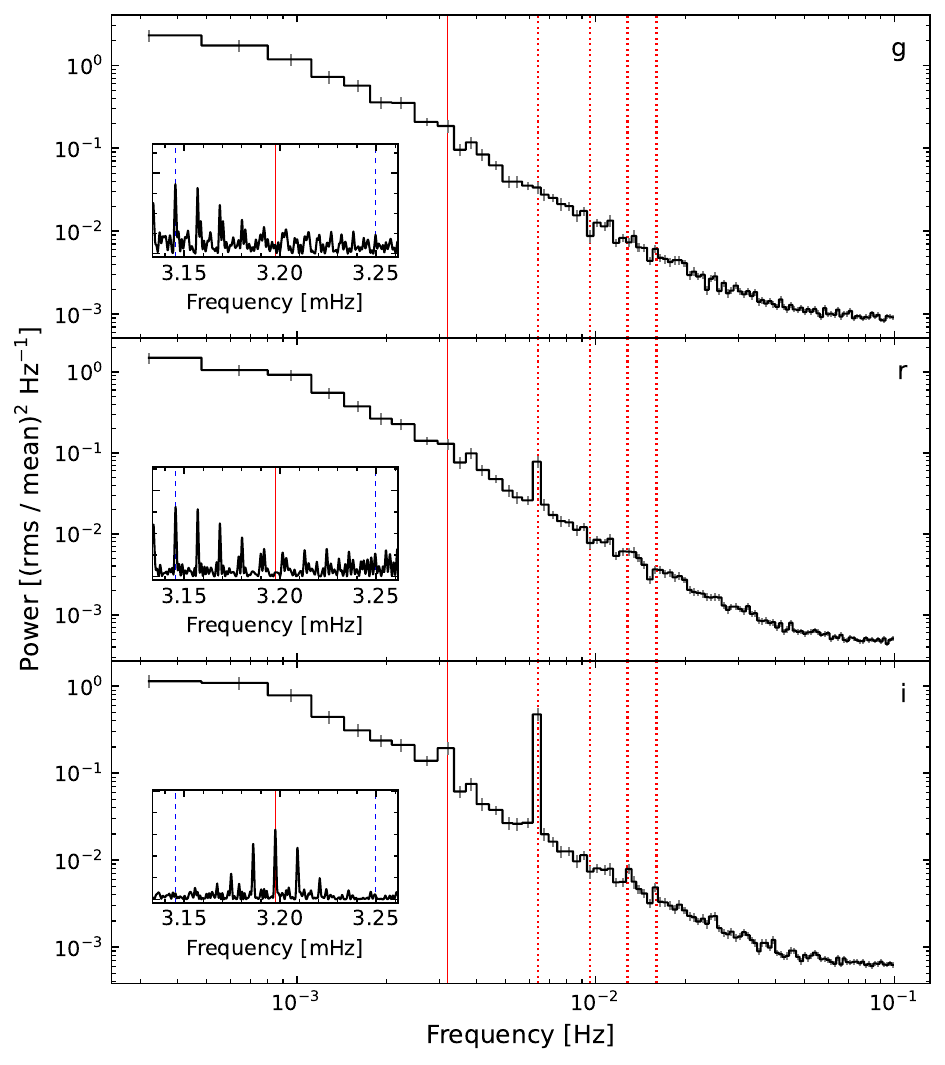}
    \caption{Power spectra of V709 Cas. Top: $g$-band; middle: $r$-band; bottom: $i$-band. The solid vertical line represents the WD spin frequency, while the dotted vertical lines represent the first four harmonics of the spin. The inset axes show the LSP for each band computed from the combined light curves from all nights. The solid vertical line in the LSP insets corresponds to the WD spin frequency, while the dashed vertical lines represent the spin--orbit sidebands.}
    \label{fig: power spectra}
\end{figure}

In Figure \ref{fig: power spectra}, we present the $g$-, $r$-, and $i$-band power spectra of V709 Cas. We also highlight the WD spin frequency (solid vertical line) and its first four harmonics (dotted vertical lines) in each spectrum. These power spectra were computed using 35 $g$- and $i$-band segments, and 39 $r$-band segments. We note that the frequency resolution of these power spectra is such that the spin and spin-orbit beat fall within the same Fourier frequency. We therefore computed the Lomb-Scargle periodograms \citep[LSPs;][]{Lomb1976, Scargle1982} for each band using the combined light curves from all nights; these LSPs, centred on the WD spin frequency, are shown in the insets of each each power spectrum.

From Figure \ref{fig: power spectra}, it can be seen that the $g$-band power spectrum shows no obvious signs of periodic variability. In contrast, the $r$-band power spectrum shows a prominent peak at the first harmonic of the WD spin frequency, which becomes even more prominent in the $i$-band power spectrum. The $i$-band power spectrum also show a weak feature at the fundamental WD spin frequency, as well as at its third and fourth harmonics. In terms of broadband variability, all three power spectra appear qualitatively similar. From the LSP insets, it can be seen that the $g$- and $r$-band LSPs show prominent features at the spin--orbit beat frequency, while the $i$-band LSP shows a prominent feature at the WD spin frequency - as reported by \cite{Irving2026}. We note that all three LSPs are heavily aliased due to the nightly observing cadence.

\subsection{Optical spectrum}\label{sec: spectrum results}

\begin{figure}
    \centering
    \includegraphics[width=\columnwidth]{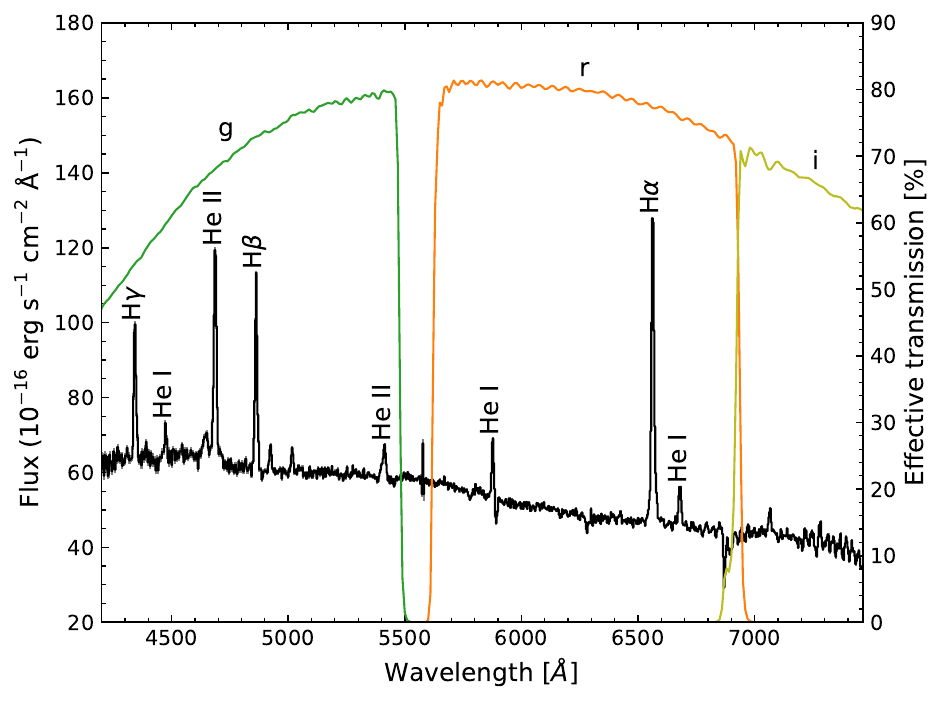}
    \caption{Optical spectrum of V709 Cas. The Balmer series (H$_\alpha$, H$_\beta$, H$_\gamma$) and He~{\sc i}/He~{\sc ii} emission lines are identified. The narrow spike near 5580~\r{A} is a residual of the O~{\sc i} night-sky line, and is not intrinsic to the source. Additionally, the broad excess at $\sim$5500~\r{A} and the absorption feature at $\sim$6680~\r{A} are both atmospheric. For reference, the effective transmission curves (detector quantum efficiency multiplied by filter transmission) of the overlapping OPTICAM filters are also shown.}
    \label{fig: spectrum}
\end{figure}

In Figure \ref{fig: spectrum}, we present the optical spectrum of V709 Cas. For reference, we also present the effective transmission curves (detector quantum efficiency multiplied by filter transmission) of the overlapping OPTICAM filters alongside this spectrum.

As can be seen from Figure \ref{fig: spectrum}, V709 Cas's optical spectrum shows prominent H$_\gamma$, He~\textsc{i}, He~\textsc{ii}, H$_\beta$, and H$_\alpha$ emission lines. It can also be seen that our $g$-band OPTICAM light curves contain contributions from the H$_\gamma$, He~\textsc{i}, He~\textsc{ii}, and H$_\beta$ emission lines, while the $r$-band contains contributions from the He~\textsc{i} and H$_\alpha$ lines. We note that this spectrum is very similar to the spectrum presented in \cite{BB2001}: the same emission lines are identified in both spectra with similar relative amplitudes, the continuum appears broadly similar (in both normalisation and shape), and we also see the narrow absorption feature on the red side of He~\textsc{i} (5875~\r{A}), which \cite{BB2001} associated with the Na~\textsc{d} doublet. However, unlike in \cite{BB2001}, we do not see broad Balmer absorption features.

\subsection{Raw coherence and lags}\label{sec: raw coh and lag results}

\begin{figure*}
    \centering
    \includegraphics[width=\textwidth]{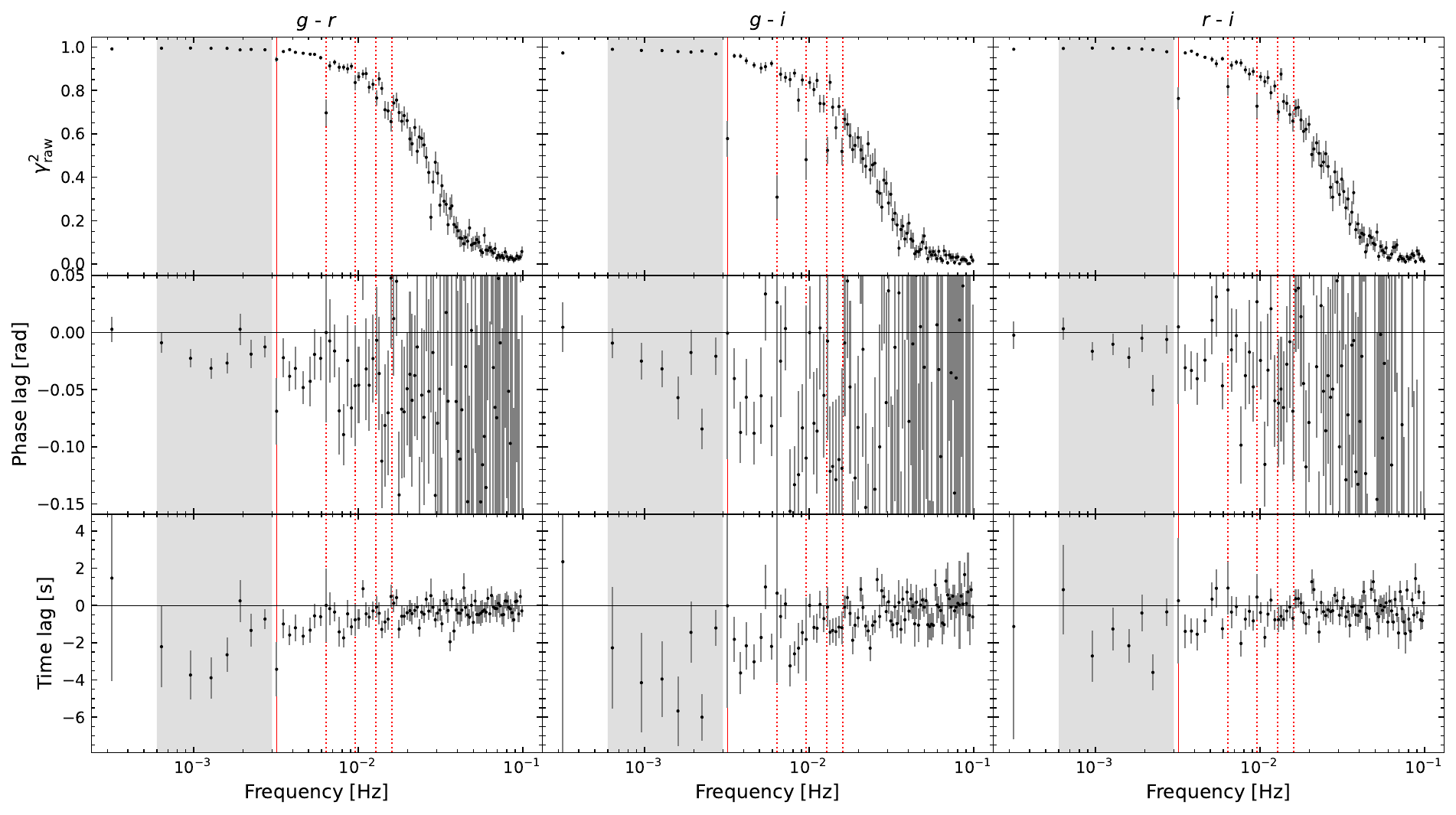}
    \caption{Raw coherence (top row), phase lags (middle row) and time lags (bottom row) of V709 Cas. The solid vertical line represents the WD spin frequency, while the dotted vertical lines represent the first four harmonics of the spin. The shaded region shows the frequency range of the detected lag. A negative lag represents the redder band lagging the bluer band; a positive lag represents the bluer band lagging the redder band.}
    \label{fig: raw coherence and lags}
\end{figure*}

The raw coherence and lags between our OPTICAM light curves are shown in Figure \ref{fig: raw coherence and lags}. In all panels, the WD spin frequency is marked with a solid vertical line, and the first four harmonics of the spin are marked with vertical dotted lines. We note that the underlying $g$--$r$, $g$--$i$, and $r$--$i$ cross spectra were computed using 35, 30, and 34 segments, respectively.

As can be seen from Figure \ref{fig: raw coherence and lags}, all three pairs of light curves show sharp drops in the raw coherence at the WD spin and its harmonics; these drops in coherence are expected given the strongly wavelength-dependent WD spin pulsations found by \cite{Irving2026}. Aside from these drops, the raw coherence is $> 0.8$ for all three spectra below approximately $10^{-2}$~Hz; above this frequency, the raw coherence falls to zero as Poisson noise dominates the underlying power spectra. As such, any features above approximately $10^{-2}$~Hz in the raw coherence or lag spectra should be considered unreliable.

Beyond the raw coherence, it can also be seen from Figure \ref{fig: raw coherence and lags} that the lags between all three pairs of light curves appear broadly negative, with redder bands lagging bluer bands. At the lowest frequency, all three pairs of light curves show a lag consistent with zero. In the $g$--$r$ and $g$--$i$ spectra, the lag can be seen to decrease with increasing frequency, reaching a minimum at $\sim$~1--2~mHz, before increasing as the frequency approaches the WD spin frequency; the frequency range of this trend roughly corresponds to the shaded grey region. In addition, there appears to be shallow local minima in the lag spectra slightly above the WD spin frequency at $\sim$~4--5~mHz; by $\sim 10^{-2}$~Hz, the lag is consistent with zero in all three spectra. In the case of the $r$--$i$ spectra, the lag is less clearly structured but appears broadly similar to the other two spectra. Quantitatively, the maximum red lags are $4 \pm 1$~s, $6 \pm 1$~s, and $4 \pm 1$~s for the $g$--$r$, $g$--$i$, and $r$--$i$ spectra, respectively. In each spectrum, the time-scale of the maximum red lag is also similar: the $g$--$r$ lag peaks at $\sim 1.28$~mHz (13.0~min), while the $g$--$i$ and $r$--$i$ lags both peak at $\sim 2.24$~mHz (7.4~min).

\subsection{Multi-spectrum fitting}\label{sec: multispectrum fitting results}

When performing the multi-spectrum fitting presented in \cite{Mendez2024}, we include fixed narrow ($Q = 10^6$) Lorentzians in the model at the WD spin frequency and its first harmonic. To minimise the number of fitted parameters, we omit higher order harmonics.

\subsubsection{$g$ and $r$ spectra}\label{sec: gxr results}

\begin{figure*}
    \centering
    \includegraphics[width=\textwidth]{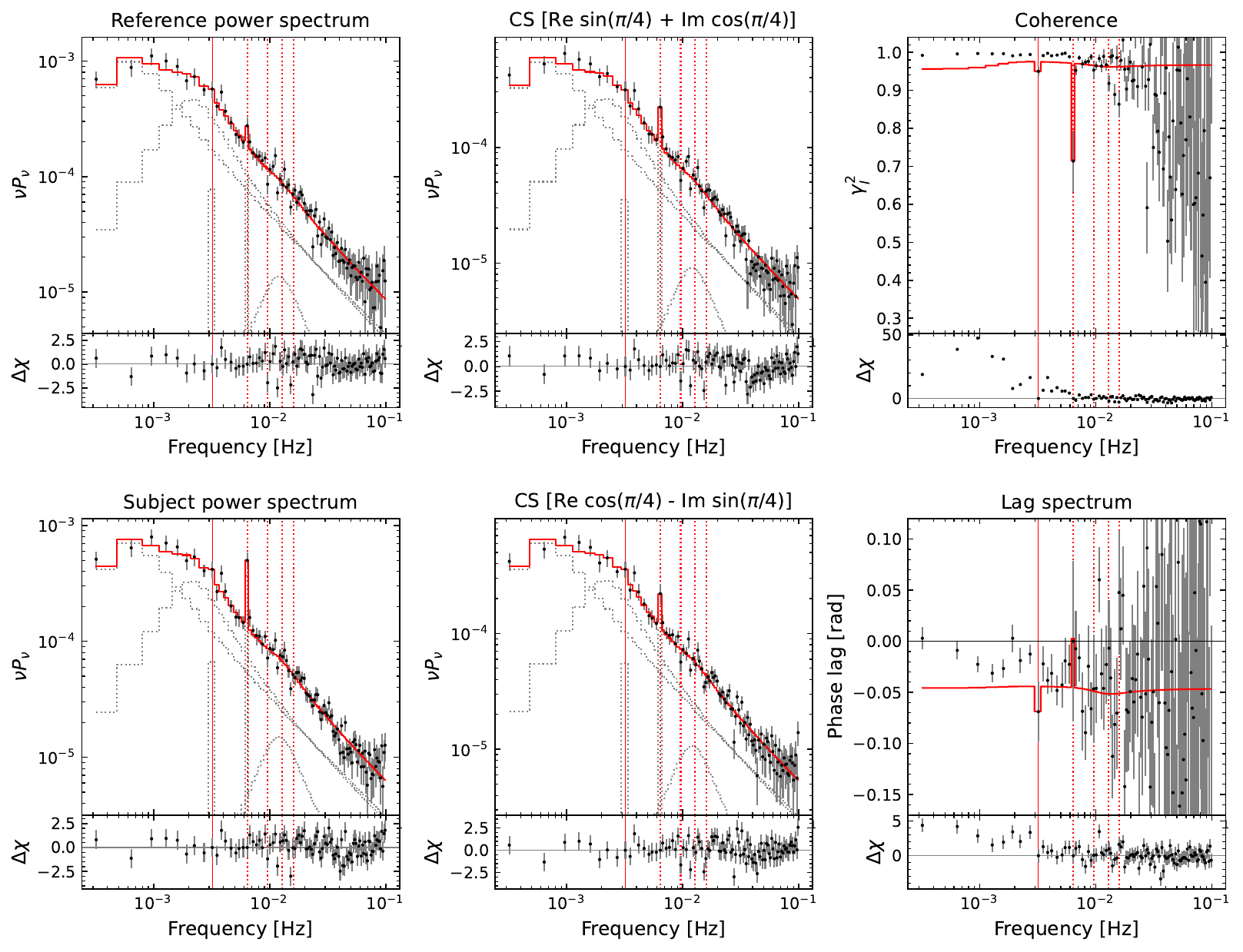}
    \caption{Joint fit to the $g$- (reference) and $r$-band (subject) cross and power spectra of V709 Cas. The fit consists of three free Lorentzians (Table \ref{tab: gxr 3 lor fit}). We note that the coherence and lag models have been derived from the joint cross and power spectra fit, and have not been fit to the measured coherence or lag spectra. The solid vertical line represents the WD spin frequency, while the dotted vertical lines represent the first four harmonics of the spin.}
    \label{fig: gxr 3 lor fit}
\end{figure*}

In Figure \ref{fig: gxr 3 lor fit}, we present the best fit to the cross and power spectra between the $g$- and $r$-bands. This fit consists of three Lorentzians in addition to the narrow Lorentzians fixed at the spin and its first harmonic ($\chi^2$/dof = 389.25/372). The parameters of this fit are presented in Table \ref{tab: gxr 3 lor fit}.

\begin{table*}
    \centering
    \caption{Three-Lorentzian fit to the cross and power spectra of the $g$- and $r$-bands shown in Figure \ref{fig: gxr 3 lor fit}.}
    \begin{tabular}{c|c|c|c|c|c|c|c}
        \hline
        Component & \multicolumn{3}{|c|}{rms [\%]} & $\nu_0$ [mHz] & $Q$ & Lag [rad] \\
         & PS$_\text{ref}$ & PS$_\text{sub}$ & CS \\
        \hline
        1 & $4.5^{+ 0.4}_{- 0.7}$ & $3.8^{+ 0.3}_{- 0.6}$ & $3.5^{+ 0.3}_{- 0.5}$ & $0.41^{+ 0.10}_{- 0.07}$ & $0.43^{+ 0.18}_{- 0.06}$ & $-0.83^{+ 0.10}_{- 0.08}$ \\
        2 & $2.8^{+ 0.9}_{- 0.7}$ & $2.3^{+ 0.7}_{- 0.6}$ & $2.1^{*}_{- 0.5}$ & $1.7 \pm 0.5$ & $0.7^{+ 0.2}_{- 0.1}$ & $-0.83^{+ 0.10}_{- 0.08}$ \\
        3 & $0.4 \pm 0.1$ & $0.4 \pm 0.1$ & $0.34^{+ 0.10}_{- 0.08}$ & $11 \pm 2$ & $1.1^{+ 0.4}_{- 0.3}$ & $-0.9 \pm 0.2$ \\
        \hline
    \end{tabular}
    \label{tab: gxr 3 lor fit}
    {
    \newline
    NOTES: $*$: unconstrained bound. All uncertainties are given to $1 \sigma$.
    }
\end{table*}

As can be seen from Figure \ref{fig: gxr 3 lor fit}, three Lorentzians clearly fails to reproduce the intrinsic coherence and lag spectrum at low frequencies. We note that an additional low-frequency Lorentzian resulted in better reproduction of the intrinsic coherence and lag spectrum (Figure \ref{fig: gxr 4 lor fit}), but its inclusion was not significant.

\subsubsection{$g$ and $i$ spectra}\label{sec: gxi results}

\begin{figure*}
    \centering
    \includegraphics[width=\textwidth]{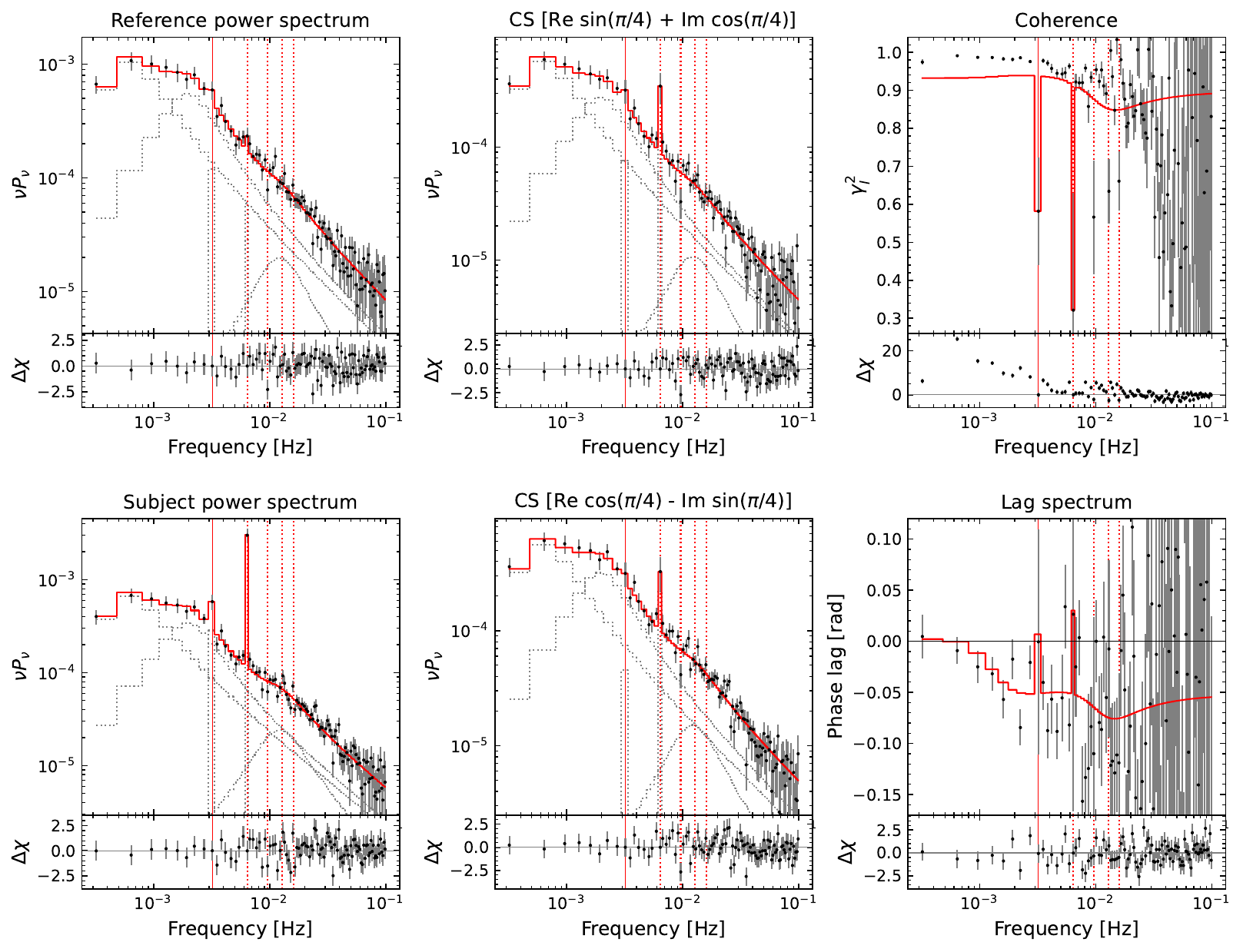}
    \caption{Joint fit to the $g$- (reference) and $i$-band (subject) cross and power spectra of V709 Cas. The fit consists of three free Lorentzians (Table \ref{tab: gxi 3 lor fit}). We note that the coherence and lag models have been derived from the joint cross and power spectra fit, and have not been fit to the measured coherence or lag spectra. The solid vertical line represents the WD spin frequency, while the dotted vertical lines represent the first four harmonics of the spin.}
    \label{fig: gxi 3 lor fit}
\end{figure*}

In Figure \ref{fig: gxi 3 lor fit}, we present the best fit to the cross and power spectra between the $g$- and $i$-bands. This fit consists of three Lorentzians in addition to the narrow Lorentzians fixed at the spin and its first harmonic ($\chi^2$/dof = 364.89/372). The parameters of this fit are presented in Table \ref{tab: gxi 3 lor fit}.

\begin{table*}
    \centering
    \caption{Three-Lorentzian fit to the cross and power spectra of the $g$- and $i$-bands shown in Figure \ref{fig: gxi 3 lor fit}.}
    \begin{tabular}{c|c|c|c|c|c|c|c}
        \hline
        Component & \multicolumn{3}{|c|}{rms [\%]} & $\nu_0$ [mHz] & $Q$ & Lag [rad] \\
         & PS$_\text{ref}$ & PS$_\text{sub}$ & CS \\
        \hline
        1 & $4.4 \pm 0.4$ & $3.5^{+ 0.3}_{- 0.4}$ & $3.3^{+ 0.2}_{- 0.3}$ & $0.44 \pm 0.04$ & $0.5^{+ 0.2}_{- 0.1}$ & $-0.78 \pm 0.09$ \\
        2 & $3.0 \pm 0.4$ & $2.3 \pm 0.4$ & $2.2 \pm 0.3$ & $1.6^{+ 0.3}_{- 0.2}$ & $0.7^{+ 0.2}_{- 0.1}$ & $-0.86^{+ 0.07}_{- 0.08}$ \\
        3 & $0.6 \pm 0.1$ & $0.6 \pm 0.1$ & $0.44^{+ 0.09}_{- 0.07}$ & $10 \pm 2$ & $0.8 \pm 0.2$ & $-1.0 \pm 0.1$ \\
        \hline
    \end{tabular}
    \label{tab: gxi 3 lor fit}
    {
    \newline
    NOTES: all uncertainties are given to $1 \sigma$.
    }
\end{table*}

Similar to Figure \ref{fig: gxr 3 lor fit}, it can be seen from Figure \ref{fig: gxi 3 lor fit} that the fit fails to reproduce the intrinsic coherence at low frequencies. In this case, however, we can reproduce the lag spectrum using only three Lorentzians, all of which are significant. As in Section \ref{sec: gxr results}, we find that we can reproduce the coherence at low frequencies by adding a fourth, non-significant Lorentzian (Figure \ref{fig: gxi 4 lor fit}).

\subsubsection{$r$ and $i$ spectra}\label{sec: rxi results}

\begin{figure*}
    \centering
    \includegraphics[width=\textwidth]{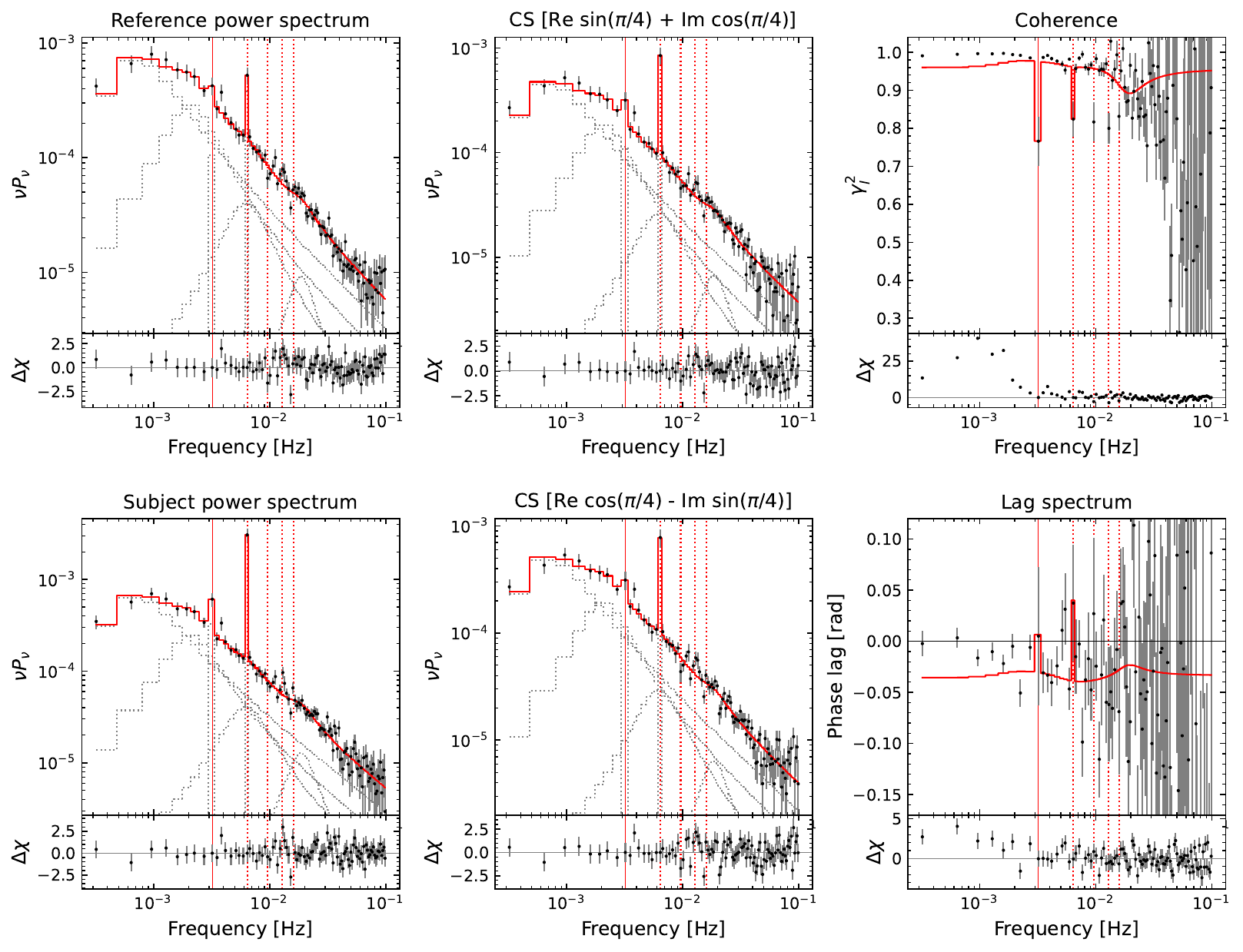}
    \caption{Joint fit to the $r$- (reference) and $i$-band (subject) cross and power spectra of V709 Cas. The fit consists of four free Lorentzians (Table \ref{tab: rxi 4 lor fit}). We note that the coherence and lag models have been derived from the joint cross and power spectra fit, and have not been fit to the measured coherence or lag spectra. The solid vertical line represents the WD spin frequency, while the dotted vertical lines represent the first four harmonics of the spin.}
    \label{fig: rxi 4 lor fit}
\end{figure*}

In Figure \ref{fig: rxi 4 lor fit}, we present the best fit to the cross and power spectra between the $r$- and $i$-bands. This fit consists of four free Lorentzians, in addition to the narrow Lorentzians fixed at the spin and its first harmonic ($\chi^2$/dof = 375.69/366). The parameters of this fit are presented in Table \ref{tab: rxi 4 lor fit}

\begin{table*}
    \centering
    \caption{Four-Lorentzian fit to the cross and power spectra of the $r$- and $i$-bands shown in Figure \ref{fig: rxi 4 lor fit}.}
    \begin{tabular}{c|c|c|c|c|c|c|c}
        \hline
        Component & \multicolumn{3}{|c|}{rms [\%]} & $\nu_0$ [mHz] & $Q$ & Lag [rad] \\
         & PS$_\text{ref}$ & PS$_\text{sub}$ & CS \\
        \hline
        1 & $3.8 \pm 0.3$ & $3.6 \pm 0.3$ & $3.1 \pm 0.2$ & $0.49 \pm 0.06$ & $0.48^{+ 0.07}_{- 0.06}$ & $-0.82^{+ 0.08}_{- 0.07}$ \\
        2 & $2.0^{+ 0.5}_{- 0.4}$ & $1.9^{+ 0.5}_{- 0.4}$ & $1.6^{+ 0.4}_{- 0.3}$ & $1.8 \pm 0.3$ & $0.9^{+ 0.3}_{- 0.2}$ & $-0.8 \pm 0.1$ \\
        3 & $0.7 \pm 0.2$ & $0.7 \pm 0.2$ & $0.6^{+ 0.2}_{- 0.1}$ & $5 \pm 1$ & $0.9^{+ 0.3}_{- 0.2}$ & $-0.8 \pm 0.2$ \\
        4 & $0.28 \pm 0.05$ & $0.33^{+ 0.06}_{- 0.05}$ & $0.24 \pm 0.04$ & $18 \pm 1$ & $1.6^{+ 0.5}_{- 0.3}$ & $-0.8 \pm 0.2$ \\
        \hline
    \end{tabular}
    \label{tab: rxi 4 lor fit}
    {
    \newline
    NOTES: all uncertainties are given to $1 \sigma$.
    }
\end{table*}

Similar to Figure \ref{fig: gxr 3 lor fit}, it can be seen from Figure \ref{fig: rxi 4 lor fit} that the fit fails to reproduce the low-frequency coherence and lag spectrum. Again, however, we find that we can only reproduce the low-frequency coherence and lag spectrum by adding a fifth, non-significant Lorentzian (Figure \ref{fig: rxi 5 lor fit}).

\section{Discussion}\label{sec: discussion}

We have discovered $\sim$~4--6~s red lags on time-scales of 7.4--13.0~min in the optical emission of V709 Cas (Section \ref{sec: raw coh and lag results}). This is the first time such lags have been detected in a magnetic CV. In addition, we also applied the multi-spectrum modelling framework of \cite{Mendez2024} to optical data of a CV for the first time (Section \ref{sec: multispectrum fitting results}). For the $g$- and $i$-bands (Section \ref{sec: gxi results}), we found that we were able to reproduce the lag spectrum following the method described in \cite{Mendez2024}. For all other bands, however, we found that a non-significant Lorentzian was required to reproduce the low-frequency intrinsic coherence and lag spectrum.

\subsection{Lags}

\subsubsection{Validating the lags}

To check whether the $\sim$4--6~s red lags found in Section \ref{sec: raw coh and lag results} are dominated by subset of nights, we would ideally perform a night-by-night analysis. Unfortunately, however, small data gaps prohibit a night-by-night analysis: within a single epoch, it is not possible to obtain $\sim 30$ segments with a sufficient frequency resolution to detect lags on 7.4--13.0~min time-scales. For fewer than 30 segments, the central limit theorem may not apply, and the resulting error bars may be unreliable \citep[e.g.,][]{Nathan2026}. To validate the lags, we therefore performed a leave-one-out consistency check. We recomputed the lag spectra using all but one epoch, recorded the minimum time lag (i.e., maximum red time lag) and its corresponding Fourier frequency, and repeated this for each epoch. We note that the resulting cross spectra were computed from as few as 24 segments, making their error bars potentially unreliable. Nonetheless, we found that the maximum red time lags changed by $\leq 1 \sigma$ for all pairs of bands. Similarly, we found that the corresponding Fourier frequencies shifted by, at most, one frequency bin. As such, we conclude that V709 Cas's $\sim$4--6~s red lags are persistent over multiple nights.

\subsubsection{Comparison with other systems}

The lags found in V709 Cas are similar to the lags seen in MV Lyr, LU Cam, and SS Cyg, both in terms of the time lags and the time-scales on which they occur \citep{Scaringi2013, Aranzana2018}. For example, we recall that \cite{Aranzana2018} found $\sim 5$~s red lags in SS Cyg on time-scales of $\sim 4.2$~min; the $\sim 4$--6~s red lags found in Section \ref{sec: raw coh and lag results} are therefore extremely similar, while the 7.4--13.0~min time-scales are only slightly longer. Given these similarities, it follows that the physical mechanism is likely similar. Below, we therefore attempt to explain V709 Cas's lags using mechanisms previously suggested for these non-magnetic CVs, as well as some relevant mechanisms proposed for XRBs.

\subsubsection{Characteristic accretion disc time-scales}\label{sec: disc time-scale discussion}

As mentioned in Section \ref{sec: intro}, \cite{Scaringi2013} suggested that the red lags in MV Lyr and LU Cam could be the result of reprocessing in the accretion disc on the thermal time-scale, or reverse (inside-out) shocks in the accretion disc - which would propagate on the thermal time-scale. Following the $\alpha$-prescription of \cite{Shakura1973}, the shortest characteristic time-scale in an accretion disc is the dynamical time-scale, $\tau_{\rm dyn}$, which is simply the Keplerian orbital period as some radius, $r$, from the compact object:
\begin{equation}
    \tau_{\rm dyn}(r) = \sqrt{\frac{4 \pi^2 r^3}{GM}},
    \label{eq: dynamical time-scale}
\end{equation}
where $G$ is the gravitational constant, and $M$ is the mass of the WD \citep[see, e.g., Chapter 5.8 of][]{Frank2002}. The thermal time-scale, $\tau_{\rm th}$, is related to the dynamical time-scale via:
\begin{equation}
    \tau_{\rm th} (r) = \frac{\tau_{\rm dyn} (r)}{\alpha (r)},
    \label{eq: thermal time-scale}
\end{equation}
where $\alpha$ is the viscosity of the disc. Since $\alpha \lesssim 1$, it follows that $\tau_{\rm th} \gtrsim \tau_{\rm dyn}$ \citep{Shakura1973}. In the case of V709 Cas, we recall that its accretion disc is truncated at the magnetospheric radius, $r_M$. Based on the break frequencies of V709 Cas's optical power spectra, \cite{Irving2026} estimated a lower limit on $r_M$ to be $1.8 \times 10^{10}$~cm. Taking this lower limit, we find:
\begin{equation*}
    \tau_{\rm dyn} (r_M) \gtrsim 1402~\text{s},
\end{equation*}
which, from the above, implies an even longer thermal-scale. Clearly, the thermal time-scale is therefore much too long to explain the $\sim$4--6~s lags found in Section \ref{sec: raw coh and lag results}; this rules-out reprocessing on the thermal time-scale and reverse (inside-out) shocks \citep{Scaringi2013}.

For completeness, we also consider the the viscous time-scale, $\tau_{\rm visc}$, which is the relevant time-scale for propagating fluctuations. As mentioned in Section \ref{sec: intro}, propagating fluctuations have been invoked to explain hard lags in XRBs \citep[e.g.,][]{Lyubarskii1997, Kotov2001, Arevalo2006, Ingram2011}; if, however, these fluctuations were to propagate in the opposite direction (i.e., inside-out), they could produce a soft lag instead. The viscous time-scale is given by:
\begin{equation}
    \tau_{\rm visc} (r) = \frac{\tau_{\rm dyn} (r)}{\alpha (r)} \Big( \frac{r}{H} \Big)^2,
\end{equation}
where $H$ is the scale-height of the disc. Assuming a thin disc, it is clear that $H \ll r$, and so $\tau_{\rm visc} \gg \tau_{\rm dyn}$ \citep{Shakura1973}. Reverse (inside-out) propagating fluctuations therefore cannot explain the $\sim$4--6~s lags found in Section \ref{sec: raw coh and lag results} either.

\subsubsection{Constraining the reprocessing region}\label{sec: emission}

Having shown in Section \ref{sec: disc time-scale discussion} that characteristic accretion disc time-scales cannot explain V709 Cas's $\sim$~4--6~s lag, we must consider alternative mechanisms. First, however, it is instructive to review the properties of V709 Cas's optical emission.

Given V709 Cas's similar optical flux during both our photometric and spectroscopic observations (Section \ref{sec: spectrum data}), we can assume that the system was likely in a similar state during both observations. We recall that our optical spectrum of V709 Cas (Figure \ref{fig: spectrum}) appears similar to the spectrum presented in \cite{BB2001}, showing Balmer (H$_\alpha$, H$_\beta$ and H$_\gamma$) and He~\textsc{i} emission lines, which can be associated directly with the accretion disc, while the strong He~\textsc{ii} emission line seen at 4686~\r{A} is characteristic of magnetically-confined accretion \citep[e.g.,][]{Saito2010}. Moreover, we also see a narrow absorption feature on the red side He~\textsc{i} (5875~\r{A}); \cite{BB2001} claimed that this absorption feature may either be interstellar in origin, or may due to the donor contributing significantly to V709 Cas's optical flux. We also note that \cite{Irving2026} found prominent spin-orbit beat pulsations in the $g$- and $r$-bands (Figure \ref{fig: power spectra}), suggesting that emission reprocessed by material fixed in the binary frame contributes significantly to these bands.

To explain V709 Cas's optical spectrum, the prominence of optical spin-orbit beat pulsations, and the $\sim$4--6~s red lags found in Section \ref{sec: raw coh and lag results}, we identify two viable mechanisms: i) reprocessing in the donor; ii) reprocessing in the bright spot, where overflowing material from the donor meets the outer-edge of the accretion disc \citep[e.g.,][]{Hassall1981}.

Considering reprocessing in the donor, V709 Cas's orbital period (5.3~hr; \citealt{Rao2026}) and WD mass ($0.88 M_\odot$; \citealt{Shaw2018}) suggest an orbital separation of $1.0 \times 10^{11}$~cm. The light travel time from the WD to the donor is therefore 3.4~s, consistent with the lags found in Section \ref{sec: raw coh and lag results}. However, this mechanism cannot explain similar lags observed in other systems. For example, in order for the $\sim$~10~s lag found in LU Cam \citep{Scaringi2013}, whose orbital period is 3.6~hr \citep{Sheets2007}, to be a result of the light travel time between the primary and the secondary, the primary would need a mass of $\sim 48$~M$_\odot$; this is clearly unphysical for a CV. Moreover, V709 Cas's optical spectrum shows no unambiguous donor contribution, casting doubt on this scenario. That said, we acknowledge that the lag mechanism in V709 Cas may be different to that of non-magnetic CVs, though it would be coincidental for the lags produced by different mechanisms to appear so similar.

Considering reprocessing in the bright spot, we still need a viable reprocessing time-scale; following \cite{Aranzana2018}, we suggest that reprocessing may occur on the recombination time-scale, $\tau_{\rm rec}$:
\begin{equation}
    \tau_{\rm rec} \approx \frac{1}{n_e \alpha_{\rm rec}},
    \label{eq: recombination time-scale}
\end{equation}
where $n_e$ is the electron density of the reprocessing region and $\alpha_{\rm rec}$ is the recombination coefficient. The recombination coefficient may be approximated by:
\begin{equation}
    \alpha_{\rm rec} \approx 1.627 \times 10^{-13} t_e^{-1/2}(1 - 1.657 \log_{10}(t_e) + 0.584 t_e^{1/3}),
\end{equation}
where $t_e \equiv 10^{-4} T_e$, and $T_e$ is the temperature of the reprocessing region \citep[][]{Hummer1963, OBrien2002}. Assuming a disc temperature of $10^4$~K and a mid-plane electron density of $10^{13}$~cm$^3$, \cite{Aranzana2018} showed that a 4~s lag could be produced by recombination in the disc at a vertical distance of $2 H$ above the mid-plane. We recall that reprocessing on the recombination time-scale has also been invoked to explain the lag in the optical emission of XRBs with respect to their X-ray emission \citep[e.g.,][]{OBrien2002}. Since the recombination time-scale does not directly depend on disc radius, and can thereby explain lags of order seconds in magnetic and non-magnetic CVs alike, we favour this mechanism over reprocessing in the donor. Moreover, recombination in the bright spot would increase the light travel time, allowing for a denser (i.e., closer to the mid-plane) and/or hotter recombination region than that proposed by \cite{Aranzana2018}.

\subsection{Multi-spectrum modelling}

In Section \ref{sec: multispectrum fitting results}, we employed the multi-spectrum modelling framework present in \cite{Mendez2024}. Within this framework, features in the coherence and lag spectra can be attributed to discrete components in the cross and power spectra. In our case, however, we found that we generally could not reproduce the low-frequency coherence or lag spectra without the addition of a non-significant low-frequency Lorentzian (see Appendix \ref{appendix: extra fits}). The exception to this is the lag spectrum between the $g$- and $i$-bands, which could be reproduced using three significant Lorentzians (Figure \ref{fig: gxi 3 lor fit}).

Since all three intrinsic coherence functions, in addition to two of the three lag spectra, required a non-significant Lorentzian to be reproduced, we suspect we might have seen better results if we had access to an improved frequency resolution. However, increasing the frequency resolution would require reducing the number of segments; we recall that our cross spectra were computed using as few as 30 segments (Section \ref{sec: raw coh and lag results})\footnote{We note that we had slightly more power spectra segments in Section \ref{sec: power spectra results} since these do not require simultaneity between bands.}. We therefore cannot substantially improve the frequency resolution while maintaining meaningful error bars on our cross and power spectra \citep[e.g.,][]{Nathan2026}. Nonetheless, we argue that the results found in Section \ref{sec: multispectrum fitting results}, and Appendix \ref{appendix: extra fits}, are encouraging. With higher time resolution and longer baseline observations than those used in this work, the multi-spectrum modelling framework presented in \cite{Mendez2024} may therefore provide a novel way to predict the intrinsic coherence and, perhaps more interestingly, lag spectra of CVs.

\subsubsection{Power spectra}

Finally, it is interesting to note that V709 Cas's optical power spectra (Figure \ref{fig: power spectra}) exhibit breaks at similar frequencies to those of ``non-magnetic'' CVs \citep[e.g.,][]{Scaringi2012, Dobrotka2015, Balman2019}. For a truncated accretion disc, the low-frequency break corresponds to the viscous frequency of the disc at the magnetospheric radius \citep[][]{Lyubarskii1997, Arevalo2006, Ingram2011, Scaringi2014}. For non-magnetic systems, the break frequency corresponds to the transition from an optically thick disc to an optically thin inner flow \citep[e.g.,][]{Scaringi2014, Balman2019}. Given these differences, there is no physical reason why the power spectra of IPs and non-magnetic CVs should break at similar frequencies. It may therefore be a coincidence, or it may indicate that the propagating fluctuations model needs to be revisited - though this is beyond the scope of this work.

\section{Conclusions}\label{sec: conclusions}

Using data from OPTICAM, we report the discovery of $\sim$~4--6~s red lags in the optical emission of the IP V709 Cas on time-scales of 7.4--13.0~min. This is the first time lags have been reported for a magnetic CV. Similar lags have been seen in the non-magnetic CVs MV Lyr and LU Cam \citep{Scaringi2013}, and SS Cyg \citep{Aranzana2018}; we therefore speculate that the physical mechanism producing these lags is also similar.

To explain V709 Cas's lags, we considered reprocessing on the thermal time-scale, reverse (inside-out) shocks, reverse (inside-out) propagating fluctuations, and reprocessing on the recombination time-scale. Since IPs possess truncated accretion discs, mechanisms occurring on characteristic accretion disc time-scales (propagating fluctuations/reprocessing on the thermal time-scale/shocks) predict lags that are orders of magnitude too large. In contrast, we showed that the light travel time between the WD and the donor is 3.4~s. While V709 Cas's lags could therefore be a result of reprocessing in the donor, we showed that this mechanism cannot explain the lags in other CVs (e.g., MV Lyr). It would therefore be coincidental for lags produced by different mechanisms to appear so similar. Of the mechanisms considered here, reprocessing in the disc on the recombination time-scale, as suggested for SS Cyg \citep{Aranzana2018}, appears the most likely. As an example, we showed that recombination in the disc's bright spot, where overflowing material from the secondary meets the outer-edge of the accretion disc, plausibly explains V709 Cas's optical spectrum, the prominence of optical spin-orbit beat pulsations \citep{Irving2026}, and the $\sim$~4--6~s lags lags found in this work. Reprocessing on the recombination time-scale also does not depend directly on disc radius, thereby explaining lags in magnetic and non-magnetic CVs alike. However, we note that the recombination time-scale is degenerate under changes in electron density and temperature, making it difficult to constrain the recombination region from a time-lag alone.

To follow-up this work, we suggest long-term multi-colour optical monitoring CVs in order to search for lags. Identifying lags in CVs is much more difficult than in XRBs for two primary reasons: i) the time-scales of variability are typically much longer in CVs \citep[e.g.,][]{Belloni2002, Revnivtsev2010}; ii) CVs are most-easily observed in the optical \citep[e.g.,][]{Warner1995}. As a result of i), long baseline observations are required. As a result of ii), obtaining simultaneous time series requires observing simultaneously with multiple independent instruments, or observing with multi-camera instruments like OPTICAM. Moreover, i) and ii) compound together: telescope time is competitive, and so obtaining long baseline simultaneous multi-band optical observations can be especially difficult. However, such observations are imperative, and characterising lags in a range of CVs is vital to be able to understand the physical mechanism(s) producing them. In the near-future, the \textit{PLATO} mission \citep{PLATO} could provide an opportunity to study CV lags with an unprecedented frequency resolution. Finally, it is interesting to note that the few CVs found to exhibit lags all exhibit red lags \citep[][]{Scaringi2013, Bruch2015, Aranzana2018}. Systematic searches for lags in CVs may therefore reveal blue lags, as seen in XRBs \citep[e.g.,][]{Kotov2001, Arevalo2006}. Blue lags may occur on different time-scales to red lags, as in XRBs, and would likely be the result of different physical mechanisms.

\section*{Acknowledgements}

We thank the referee for their comments and suggestions, which have improved the quality of this manuscript.

This work is based upon observations carried out at the Observatorio Astron\'omico Nacional on the Sierra San Pedro M\'artir, Baja California, M\'exico.

Z.A.I gratefully acknowledges support from the UK Research and Innovation's Science and Technology Facilities Council (STFC) grant ST/X508767/1. D.d.M acknowledges support from the INAF. N.C.S acknowledge support from the Science and Technology Facilities Council (STFC) grant ST/X001121/1.

The authors acknowledge the use of the \texttt{astropy} \citep{astropy}, \texttt{extinction} \citep{Barbary2016}, \texttt{IRAF} \citep{Tody1986, Tody1993}, \texttt{LMFIT} \citep{LMFIT}, \texttt{matplotlib} \citep{Hunter2007}, \texttt{numpy} \citep{Harris2020}, and \texttt{scipy} \citep{SciPy2020} software.

\section*{Data Availability}

The data analysed in this work will be made available upon request. We note that a public reduction pipeline for OPTICAM is currently in development.



\bibliographystyle{mnras}
\bibliography{example} 

@ARTICLE{Mendez2024,
       author = {{M{\'e}ndez}, Mariano and {Peirano}, Valentina and {Garc{\'\i}a}, Federico and {Belloni}, Tomaso and {Altamirano}, Diego and {Alabarta}, Kevin},
        title = "{Unveiling hidden variability components in accreting X-ray binaries using both the Fourier power and cross-spectra}",
      journal = {\mnras},
         year = 2024,
        month = jan,
       volume = {527},
       number = {3},
        pages = {9405-9430},
          doi = {10.1093/mnras/stad3786},
archivePrefix = {arXiv},
       eprint = {2312.03476},
 primaryClass = {astro-ph.HE},
       adsurl = {https://ui.adsabs.harvard.edu/abs/2024MNRAS.527.9405M}
}

@ARTICLE{Irving2026,
       author = {{Irving}, Z.~A. and {Castro Segura}, N. and {Altamirano}, D. and {Scaringi}, S. and {Veresvarska}, M. and {Vincentelli}, F. and {de Martino}, D. and {Buckley}, D.~A.~H. and {Castro}, A. and {Michel}, R.},
        title = "{OPTICAM reveals hints of cyclotron emission from the intermediate polar V709 Cas}",
      journal = {\mnras},
         year = 2026,
        month = jan,
       volume = {545},
       number = {3},
          eid = {staf2177},
        pages = {staf2177},
          doi = {10.1093/mnras/staf2177},
       adsurl = {https://ui.adsabs.harvard.edu/abs/2026MNRAS.545f2177I}
}

@ARTICLE{Vaughan1997,
       author = {{Vaughan}, Brian A. and {Nowak}, Michael A.},
        title = "{X-Ray Variability Coherence: How to Compute It, What It Means, and How It Constrains Models of GX 339-4 and Cygnus X-1}",
      journal = {\apjl},
         year = 1997,
        month = jan,
       volume = {474},
       number = {1},
        pages = {L43-L46},
          doi = {10.1086/310430},
archivePrefix = {arXiv},
       eprint = {astro-ph/9610257},
 primaryClass = {astro-ph},
       adsurl = {https://ui.adsabs.harvard.edu/abs/1997ApJ...474L..43V}
}

@ARTICLE{Scaringi2013,
       author = {{Scaringi}, S. and {K{\"o}rding}, E. and {Groot}, P.~J. and {Uttley}, P. and {Marsh}, T. and {Knigge}, C. and {Maccarone}, T. and {Dhillon}, V.~S.},
        title = "{Discovery of Fourier-dependent time lags in cataclysmic variables}",
      journal = {\mnras},
         year = 2013,
        month = may,
       volume = {431},
       number = {3},
        pages = {2535-2541},
          doi = {10.1093/mnras/stt347},
archivePrefix = {arXiv},
       eprint = {1302.5422},
 primaryClass = {astro-ph.GA},
       adsurl = {https://ui.adsabs.harvard.edu/abs/2013MNRAS.431.2535S}
}

@ARTICLE{Aranzana2018,
       author = {{Aranzana}, E. and {Scaringi}, S. and {K{\"o}rding}, E. and {Dhillon}, V.~S. and {Coppejans}, D.~L.},
        title = "{Fourier time lags in the dwarf nova SS Cygni}",
      journal = {\mnras},
         year = 2018,
        month = dec,
       volume = {481},
       number = {2},
        pages = {2140-2147},
          doi = {10.1093/mnras/sty2367},
archivePrefix = {arXiv},
       eprint = {1808.09253},
 primaryClass = {astro-ph.SR},
       adsurl = {https://ui.adsabs.harvard.edu/abs/2018MNRAS.481.2140A}
}

@ARTICLE{Bruch2015,
       author = {{Bruch}, Albert},
        title = "{Time lags of the flickering in cataclysmic variables as a function of wavelength}",
      journal = {\aap},
         year = 2015,
        month = jul,
       volume = {579},
          eid = {A50},
        pages = {A50},
          doi = {10.1051/0004-6361/201425393},
archivePrefix = {arXiv},
       eprint = {1503.02982},
 primaryClass = {astro-ph.SR},
       adsurl = {https://ui.adsabs.harvard.edu/abs/2015A&A...579A..50B}
}

@ARTICLE{Dhillon2007,
       author = {{Dhillon}, V.~S. and {Marsh}, T.~R. and {Stevenson}, M.~J. and {Atkinson}, D.~C. and {Kerry}, P. and {Peacocke}, P.~T. and {Vick}, A.~J.~A. and {Beard}, S.~M. and {Ives}, D.~J. and {Lunney}, D.~W. and {McLay}, S.~A. and {Tierney}, C.~J. and {Kelly}, J. and {Littlefair}, S.~P. and {Nicholson}, R. and {Pashley}, R. and {Harlaftis}, E.~T. and {O'Brien}, K.},
        title = "{ULTRACAM: an ultrafast, triple-beam CCD camera for high-speed astrophysics}",
      journal = {\mnras},
         year = 2007,
        month = jul,
       volume = {378},
       number = {3},
        pages = {825-840},
          doi = {10.1111/j.1365-2966.2007.11881.x},
archivePrefix = {arXiv},
       eprint = {0704.2557},
 primaryClass = {astro-ph},
       adsurl = {https://ui.adsabs.harvard.edu/abs/2007MNRAS.378..825D}
}

@BOOK{Warner1995,
       author = {{Warner}, Brian},
        title = "{Cataclysmic variable stars}",
         year = 1995,
       volume = {28},
       series = "", 
       adsurl = {https://ui.adsabs.harvard.edu/abs/1995cvs..book.....W}
}

@ARTICLE{Warner2004,
       author = {{Warner}, Brian},
        title = "{Rapid Oscillations in Cataclysmic Variables}",
      journal = {\pasp},
         year = 2004,
        month = feb,
       volume = {116},
       number = {816},
        pages = {115-132},
          doi = {10.1086/381742},
archivePrefix = {arXiv},
       eprint = {astro-ph/0312182},
 primaryClass = {astro-ph},
       adsurl = {https://ui.adsabs.harvard.edu/abs/2004PASP..116..115W}
}

@ARTICLE{Patterson1994,
       author = {{Patterson}, Joseph},
        title = "{The DQ Herculis Stars}",
      journal = {\pasp},
         year = 1994,
        month = mar,
       volume = {106},
        pages = {209},
          doi = {10.1086/133375},
       adsurl = {https://ui.adsabs.harvard.edu/abs/1994PASP..106..209P}
}

@ARTICLE{Lyubarskii1997,
       author = {{Lyubarskii}, Yu. E.},
        title = "{Flicker noise in accretion discs}",
      journal = {\mnras},
         year = 1997,
        month = dec,
       volume = {292},
       number = {3},
        pages = {679-685},
          doi = {10.1093/mnras/292.3.679},
       adsurl = {https://ui.adsabs.harvard.edu/abs/1997MNRAS.292..679L}
}

@ARTICLE{Arevalo2006,
       author = {{Ar{\'e}valo}, P. and {Uttley}, P.},
        title = "{Investigating a fluctuating-accretion model for the spectral-timing properties of accreting black hole systems}",
      journal = {\mnras},
         year = 2006,
        month = apr,
       volume = {367},
       number = {2},
        pages = {801-814},
          doi = {10.1111/j.1365-2966.2006.09989.x},
archivePrefix = {arXiv},
       eprint = {astro-ph/0512394},
 primaryClass = {astro-ph},
       adsurl = {https://ui.adsabs.harvard.edu/abs/2006MNRAS.367..801A}
}

@ARTICLE{Castro2019,
       author = {{Castro}, A. and {Altamirano}, D. and {Michel}, R. and {Gandhi}, P. and {Hern{\'a}ndez Santisteban}, J.~V. and {Echevarr{\'\i}a}, J. and {Tejada}, C. and {Knigge}, C. and {Sierra}, G. and {Colorado}, E. and {Hern{\'a}ndez-Landa}, J. and {Whiter}, D. and {Middleton}, M. and {Garc{\'\i}a}, B. and {Guisa}, G. and {CastroSegura}, N.},
        title = "{OPTICAM: A Triple-Camera Optical System Designed to Explore the Fastest Timescales in Astronomy}",
      journal = {\rmxaa},
         year = 2019,
        month = oct,
       volume = {55},
        pages = {363-376},
          doi = {10.22201/ia.01851101p.2019.55.02.20},
archivePrefix = {arXiv},
       eprint = {1908.05785},
 primaryClass = {astro-ph.IM},
       adsurl = {https://ui.adsabs.harvard.edu/abs/2019RMxAA..55..363C}
}

@ARTICLE{Nathan2026,
       author = {{Nathan}, Edward J.~R. and {Ingram}, Adam and {Huppenkothen}, Daniela and {Bachetti}, Matteo and {Garc{\'\i}a}, Javier A.},
        title = "{The statistical properties of the cross spectrum}",
      journal = {RAS Techniques and Instruments},
         year = 2026,
        month = jan,
       volume = {5},
          eid = {rzag010},
        pages = {rzag010},
          doi = {10.1093/rasti/rzag010},
archivePrefix = {arXiv},
       eprint = {2603.02094},
 primaryClass = {astro-ph.IM},
       adsurl = {https://ui.adsabs.harvard.edu/abs/2026RASTI...5ag010N}
}

@ARTICLE{Shakura1973,
       author = {{Shakura}, N.~I. and {Sunyaev}, R.~A.},
        title = "{Black holes in binary systems. Observational appearance.}",
      journal = {\aap},
         year = 1973,
        month = jan,
       volume = {24},
        pages = {337-355},
       adsurl = {https://ui.adsabs.harvard.edu/abs/1973A&A....24..337S}
}

@ARTICLE{Shaw2018,
       author = {{Shaw}, A.~W. and {Heinke}, C.~O. and {Mukai}, K. and {Sivakoff}, G.~R. and {Tomsick}, J.~A. and {Rana}, V.},
        title = "{Measuring the masses of intermediate polars with NuSTAR: V709 Cas, NY Lup, and V1223 Sgr}",
      journal = {\mnras},
         year = 2018,
        month = may,
       volume = {476},
       number = {1},
        pages = {554-561},
          doi = {10.1093/mnras/sty246},
archivePrefix = {arXiv},
       eprint = {1801.08508},
 primaryClass = {astro-ph.HE},
       adsurl = {https://ui.adsabs.harvard.edu/abs/2018MNRAS.476..554S}
}

@ARTICLE{Kotov2001,
       author = {{Kotov}, O. and {Churazov}, E. and {Gilfanov}, M.},
        title = "{On the X-ray time-lags in the black hole candidates}",
      journal = {\mnras},
         year = 2001,
        month = nov,
       volume = {327},
       number = {3},
        pages = {799-807},
          doi = {10.1046/j.1365-8711.2001.04769.x},
archivePrefix = {arXiv},
       eprint = {astro-ph/0103115},
 primaryClass = {astro-ph},
       adsurl = {https://ui.adsabs.harvard.edu/abs/2001MNRAS.327..799K}
}

@ARTICLE{Kaaret1999,
       author = {{Kaaret}, Philip and {Piraino}, Santina and {Ford}, Eric C. and {Santangelo}, Andrea},
        title = "{Discovery of Microsecond Soft Lags in the X-Ray Emission of the Atoll Source 4U 1636-536}",
      journal = {\apjl},
         year = 1999,
        month = mar,
       volume = {514},
       number = {1},
        pages = {L31-L33},
          doi = {10.1086/311941},
archivePrefix = {arXiv},
       eprint = {astro-ph/9901349},
 primaryClass = {astro-ph},
       adsurl = {https://ui.adsabs.harvard.edu/abs/1999ApJ...514L..31K}
}

@ARTICLE{Vaughan1998,
       author = {{Vaughan}, B.~A. and {van der Klis}, M. and {M{\'e}ndez}, M. and {van Paradijs}, J. and {Wijnands}, R.~A.~D. and {Lewin}, W.~H.~G. and {Lamb}, F.~K. and {Psaltis}, D. and {Kuulkers}, E. and {Oosterbroek}, T.},
        title = "{Erratum: Discovery of Microsecond Time Lags in Kilohertz QPOs}",
      journal = {\apjl},
         year = 1998,
        month = dec,
       volume = {509},
       number = {2},
        pages = {L145-L145},
          doi = {10.1086/311785},
       adsurl = {https://ui.adsabs.harvard.edu/abs/1998ApJ...509L.145V}
}

@ARTICLE{Fabian2009,
       author = {{Fabian}, A.~C. and {Zoghbi}, A. and {Ross}, R.~R. and {Uttley}, P. and {Gallo}, L.~C. and {Brandt}, W.~N. and {Blustin}, A.~J. and {Boller}, T. and {Caballero-Garcia}, M.~D. and {Larsson}, J. and {Miller}, J.~M. and {Miniutti}, G. and {Ponti}, G. and {Reis}, R.~C. and {Reynolds}, C.~S. and {Tanaka}, Y. and {Young}, A.~J.},
        title = "{Broad line emission from iron K- and L-shell transitions in the active galaxy 1H0707-495}",
      journal = {\nat},
         year = 2009,
        month = may,
       volume = {459},
       number = {7246},
        pages = {540-542},
          doi = {10.1038/nature08007},
       adsurl = {https://ui.adsabs.harvard.edu/abs/2009Natur.459..540F}
}

@ARTICLE{Zoghbi2010,
       author = {{Zoghbi}, A. and {Fabian}, A.~C. and {Uttley}, P. and {Miniutti}, G. and {Gallo}, L.~C. and {Reynolds}, C.~S. and {Miller}, J.~M. and {Ponti}, G.},
        title = "{Broad iron L line and X-ray reverberation in 1H0707-495}",
      journal = {\mnras},
         year = 2010,
        month = feb,
       volume = {401},
       number = {4},
        pages = {2419-2432},
          doi = {10.1111/j.1365-2966.2009.15816.x},
archivePrefix = {arXiv},
       eprint = {0910.0367},
 primaryClass = {astro-ph.HE},
       adsurl = {https://ui.adsabs.harvard.edu/abs/2010MNRAS.401.2419Z}
}

@ARTICLE{OBrien2002,
       author = {{O'Brien}, K. and {Horne}, Keith and {Hynes}, R.~I. and {Chen}, W. and {Haswell}, C.~A. and {Still}, M.~D.},
        title = "{Echoes in X-ray binaries}",
      journal = {\mnras},
         year = 2002,
        month = aug,
       volume = {334},
       number = {2},
        pages = {426-434},
          doi = {10.1046/j.1365-8711.2002.05530.x},
archivePrefix = {arXiv},
       eprint = {astro-ph/0204018},
 primaryClass = {astro-ph},
       adsurl = {https://ui.adsabs.harvard.edu/abs/2002MNRAS.334..426O}
}

@ARTICLE{Belloni2002,
       author = {{Belloni}, Tomaso and {Psaltis}, Dimitrios and {van der Klis}, Michiel},
        title = "{A Unified Description of the Timing Features of Accreting X-Ray Binaries}",
      journal = {\apj},
         year = 2002,
        month = jun,
       volume = {572},
       number = {1},
        pages = {392-406},
          doi = {10.1086/340290},
archivePrefix = {arXiv},
       eprint = {astro-ph/0202213},
 primaryClass = {astro-ph},
       adsurl = {https://ui.adsabs.harvard.edu/abs/2002ApJ...572..392B}
}

@ARTICLE{Hummer1963,
       author = {{Hummer}, D.~G. and {Seaton}, M.~J.},
        title = "{The ionization structure of planetary nebulae, I. Pure hydrogen nebulae}",
      journal = {\mnras},
         year = 1963,
        month = jan,
       volume = {125},
        pages = {437},
          doi = {10.1093/mnras/125.5.437},
       adsurl = {https://ui.adsabs.harvard.edu/abs/1963MNRAS.125..437H}
}

@ARTICLE{Uttley2014,
       author = {{Uttley}, P. and {Cackett}, E.~M. and {Fabian}, A.~C. and {Kara}, E. and {Wilkins}, D.~R.},
        title = "{X-ray reverberation around accreting black holes}",
      journal = {\aapr},
         year = 2014,
        month = aug,
       volume = {22},
          eid = {72},
        pages = {72},
          doi = {10.1007/s00159-014-0072-0},
archivePrefix = {arXiv},
       eprint = {1405.6575},
 primaryClass = {astro-ph.HE},
       adsurl = {https://ui.adsabs.harvard.edu/abs/2014A&ARv..22...72U}
}

@ARTICLE{Papadakis2001,
       author = {{Papadakis}, I.~E. and {Nandra}, K. and {Kazanas}, D.},
        title = "{Frequency-dependent Time Lags in the X-Ray Emission of the Seyfert Galaxy NGC 7469}",
      journal = {\apjl},
         year = 2001,
        month = jun,
       volume = {554},
       number = {2},
        pages = {L133-L137},
          doi = {10.1086/321722},
archivePrefix = {arXiv},
       eprint = {astro-ph/0106175},
 primaryClass = {astro-ph},
       adsurl = {https://ui.adsabs.harvard.edu/abs/2001ApJ...554L.133P}
}

@ARTICLE{Cropper1990,
       author = {{Cropper}, Mark},
        title = "{The Polars}",
      journal = {\ssr},
         year = 1990,
        month = dec,
       volume = {54},
       number = {3-4},
        pages = {195-295},
          doi = {10.1007/BF00177799},
       adsurl = {https://ui.adsabs.harvard.edu/abs/1990SSRv...54..195C}
}

@ARTICLE{Patterson1995,
       author = {{Patterson}, Joseph and {Skillman}, David R. and {Thorstensen}, John and {Hellier}, Coel},
        title = "{The Remarkable Eclipsing Asynchronous AM Herculis Binary RX J19402-1025}",
      journal = {\pasp},
         year = 1995,
        month = apr,
       volume = {107},
        pages = {307},
          doi = {10.1086/133556},
       adsurl = {https://ui.adsabs.harvard.edu/abs/1995PASP..107..307P}
}

@ARTICLE{Rea2017,
       author = {{Rea}, N. and {Coti Zelati}, F. and {Esposito}, P. and {D'Avanzo}, P. and {de Martino}, D. and {Israel}, G.~L. and {Torres}, D.~F. and {Campana}, S. and {Belloni}, T.~M. and {Papitto}, A. and {Masetti}, N. and {Carrasco}, L. and {Possenti}, A. and {Wieringa}, M. and {Wilhelmi}, E. De O{\~n}a and {Li}, J. and {Bozzo}, E. and {Ferrigno}, C. and {Linares}, M. and {Tauris}, T.~M. and {Hernanz}, M. and {Ribas}, I. and {Monelli}, M. and {Borghese}, A. and {Baglio}, M.~C. and {Casares}, J.},
        title = "{Multiband study of RX J0838-2827 and XMM J083850.4-282759: a new asynchronous magnetic cataclysmic variable and a candidate transitional millisecond pulsar}",
      journal = {\mnras},
         year = 2017,
        month = nov,
       volume = {471},
       number = {3},
        pages = {2902-2916},
          doi = {10.1093/mnras/stx1560},
archivePrefix = {arXiv},
       eprint = {1611.04194},
 primaryClass = {astro-ph.HE},
       adsurl = {https://ui.adsabs.harvard.edu/abs/2017MNRAS.471.2902R}
}

@ARTICLE{Tovmassian2017,
       author = {{Tovmassian}, G. and {Gonz{\'a}lez-Buitrago}, D. and {Thorstensen}, J. and {Kotze}, E. and {Breytenbach}, H. and {Schwope}, A. and {Bernardini}, F. and {Zharikov}, S.~V. and {Hernandez}, M.~S. and {Buckley}, D.~A.~H. and {de Miguel}, E. and {Hambsch}, F.-J. and {Myers}, G. and {Goff}, W. and {Cejudo}, D. and {Starkey}, D. and {Campbell}, T. and {Ulowetz}, J. and {Stein}, W. and {Nelson}, P. and {Reichart}, D.~E. and {Haislip}, J.~B. and {Ivarsen}, K.~M. and {LaCluyze}, A.~P. and {Moore}, J.~P. and {Miroshnichenko}, A.~S.},
        title = "{IGR J19552+0044: A new asynchronous short period polar. Filling the gap between intermediate and ordinary polars}",
      journal = {\aap},
         year = 2017,
        month = dec,
       volume = {608},
          eid = {A36},
        pages = {A36},
          doi = {10.1051/0004-6361/201731323},
archivePrefix = {arXiv},
       eprint = {1710.02126},
 primaryClass = {astro-ph.SR},
       adsurl = {https://ui.adsabs.harvard.edu/abs/2017A&A...608A..36T}
}

@ARTICLE{Rao2026,
       author = {{Rao}, Srinivas M. and {Pandey}, Jeewan C. and {Rawat}, Nikita and {Joshi}, Arti and {Singh}, Ajay Kumar},
        title = "{Long-term optical photometry of V709 Cas using TESS: Refined periods and accretion geometry}",
      journal = {\na},
         year = 2026,
        month = jan,
       volume = {122},
          eid = {102481},
        pages = {102481},
          doi = {10.1016/j.newast.2025.102481},
archivePrefix = {arXiv},
       eprint = {2507.19441},
 primaryClass = {astro-ph.SR},
       adsurl = {https://ui.adsabs.harvard.edu/abs/2026NewA..12202481R}
}

@ARTICLE{Motch1996,
       author = {{Motch}, C. and {Haberl}, F. and {Guillout}, P. and {Pakull}, M. and {Reinsch}, K. and {Krautter}, J.},
        title = "{New cataclysmic variables from the ROSAT All-Sky Survey.}",
      journal = {\aap},
         year = 1996,
        month = mar,
       volume = {307},
        pages = {459-469},
       adsurl = {https://ui.adsabs.harvard.edu/abs/1996A&A...307..459M}
}

@ARTICLE{Haberl1995,
       author = {{Haberl}, F. and {Motch}, C.},
        title = "{New intermediate polars discovered in the ROSAT survey: two spectrally distinct classes.}",
      journal = {\aap},
         year = 1995,
        month = may,
       volume = {297},
        pages = {L37},
       adsurl = {https://ui.adsabs.harvard.edu/abs/1995A&A...297L..37H}
}

@ARTICLE{Sheets2007,
       author = {{Sheets}, Holly A. and {Thorstensen}, John R. and {Peters}, Christopher J. and {Kapusta}, Ann B. and {Taylor}, Cynthia J.},
        title = "{Spectroscopy of Nine Cataclysmic Variable Stars}",
      journal = {\pasp},
         year = 2007,
        month = may,
       volume = {119},
       number = {855},
        pages = {494-507},
          doi = {10.1086/518698},
archivePrefix = {arXiv},
       eprint = {0704.0948},
 primaryClass = {astro-ph},
       adsurl = {https://ui.adsabs.harvard.edu/abs/2007PASP..119..494S}
}

@ARTICLE{BB2001,
       author = {{Bonnet-Bidaud}, J.~M. and {Mouchet}, M. and {de Martino}, D. and {Matt}, G. and {Motch}, C.},
        title = "{The white dwarf revealed in the intermediate polar V709 Cassiopeiae}",
      journal = {\aap},
         year = 2001,
        month = aug,
       volume = {374},
        pages = {1003-1008},
          doi = {10.1051/0004-6361:20010756},
archivePrefix = {arXiv},
       eprint = {astro-ph/0106021},
 primaryClass = {astro-ph},
       adsurl = {https://ui.adsabs.harvard.edu/abs/2001A&A...374.1003B}
}

@INPROCEEDINGS{vanderKlis1989,
       author = {{van der Klis}, M.},
        title = "{Fourier techniques in X-ray timing}",
    booktitle = {Timing Neutron Stars},
         year = 1989,
       editor = {{{\"O}gelman}, H. and {van den Heuvel}, E.~P.~J.},
       series = {NATO Advanced Study Institute (ASI) Series C},
       volume = {262},
        month = jan,
        pages = {27},
          doi = {10.1007/978-94-009-2273-0_3},
       adsurl = {https://ui.adsabs.harvard.edu/abs/1989ASIC..262...27V}
}

@ARTICLE{Reig2000,
       author = {{Reig}, P. and {Belloni}, T. and {van der Klis}, M. and {M{\'e}ndez}, M. and {Kylafis}, N.~D. and {Ford}, E.~C.},
        title = "{Phase Lag Variability Associated with the 0.5-10 HZ Quasi-Periodic Oscillations in GRS 1915+105}",
      journal = {\apj},
         year = 2000,
        month = oct,
       volume = {541},
       number = {2},
        pages = {883-888},
          doi = {10.1086/309469},
       adsurl = {https://ui.adsabs.harvard.edu/abs/2000ApJ...541..883R}
}

@ARTICLE{Hassall1981,
       author = {{Hassall}, B.~J.~M. and {Pringle}, J.~E. and {Ward}, M.~J. and {Whelan}, J.~A.~J. and {Mayo}, S.~K. and {Echevarria}, J. and {Jones}, D.~H.~P. and {Wallis}, R.~E. and {Allen}, D.~A. and {Hyland}, A.~R.},
        title = "{Observations and models of H 2252-035.}",
      journal = {\mnras},
         year = 1981,
        month = oct,
       volume = {197},
        pages = {275-286},
          doi = {10.1093/mnras/197.2.275},
       adsurl = {https://ui.adsabs.harvard.edu/abs/1981MNRAS.197..275H}
}

@ARTICLE{Kozhevnikov2001,
       author = {{Kozhevnikov}, V.~P.},
        title = "{Detection of optical oscillations of the intermediate polar <ASTROBJ>V709 Cassiopeae</ASTROBJ> <ASTROBJ>(RX J0028.8+5917)</ASTROBJ>}",
      journal = {\aap},
         year = 2001,
        month = feb,
       volume = {366},
        pages = {891-897},
          doi = {10.1051/0004-6361:20000259},
       adsurl = {https://ui.adsabs.harvard.edu/abs/2001A&A...366..891K}
}

@ARTICLE{Tamburini2009,
       author = {{Tamburini}, F. and {de Martino}, D. and {Bianchini}, A.},
        title = "{Analysis of the white-light flickering of the intermediate polar V709 Cassiopeiae with wavelets and Hurst analysis}",
      journal = {\aap},
         year = 2009,
        month = jul,
       volume = {502},
       number = {1},
        pages = {1-5},
          doi = {10.1051/0004-6361/200911656},
archivePrefix = {arXiv},
       eprint = {0901.1767},
 primaryClass = {astro-ph.SR},
       adsurl = {https://ui.adsabs.harvard.edu/abs/2009A&A...502....1T}
}

@ARTICLE{Hric2014,
       author = {{Hric}, L. and {Breus}, V. and {Katysheva}, N.~A. and {Shugarov}, S. Yu. and {Dubovsk{\'y}}, P.},
        title = "{The new period of the intermediate polar V709 Cas}",
      journal = {Astronomische Nachrichten},
         year = 2014,
        month = jan,
       volume = {335},
       number = {4},
        pages = {362},
          doi = {10.1002/asna.201312044},
       adsurl = {https://ui.adsabs.harvard.edu/abs/2014AN....335..362H}
}

@ARTICLE{Norton1999,
       author = {{Norton}, A.~J. and {Beardmore}, A.~P. and {Allan}, A. and {Hellier}, C.},
        title = "{YY Draconis and V709 Cassiopeiae: two intermediate polars with weak magnetic fields}",
      journal = {\aap},
         year = 1999,
        month = jul,
       volume = {347},
        pages = {203-211},
          doi = {10.48550/arXiv.astro-ph/9811310},
archivePrefix = {arXiv},
       eprint = {astro-ph/9811310},
 primaryClass = {astro-ph},
       adsurl = {https://ui.adsabs.harvard.edu/abs/1999A&A...347..203N}
}

@ARTICLE{DDM2001,
       author = {{de Martino}, D. and {Matt}, G. and {Mukai}, K. and {Belloni}, T. and {Bonnet-Bidaud}, J.~M. and {Chiappetti}, L. and {G{\"a}nsicke}, B.~T. and {Haberl}, F. and {Mouchet}, M.},
        title = "{The X-ray emission of the intermediate polar V 709 Cas}",
      journal = {\aap},
         year = 2001,
        month = oct,
       volume = {377},
        pages = {499-511},
          doi = {10.1051/0004-6361:20011059},
archivePrefix = {arXiv},
       eprint = {astro-ph/0107480},
 primaryClass = {astro-ph},
       adsurl = {https://ui.adsabs.harvard.edu/abs/2001A&A...377..499D}
}

@ARTICLE{Mukai2015,
       author = {{Mukai}, K. and {Rana}, V. and {Bernardini}, F. and {de Martino}, D.},
        title = "{Unambiguous Detection of Reflection in Magnetic Cataclysmic Variables: Joint NuSTAR-XMM-Newton Observations of Three Intermediate Polars}",
      journal = {\apjl},
         year = 2015,
        month = jul,
       volume = {807},
       number = {2},
          eid = {L30},
        pages = {L30},
          doi = {10.1088/2041-8205/807/2/L30},
archivePrefix = {arXiv},
       eprint = {1506.07213},
 primaryClass = {astro-ph.HE},
       adsurl = {https://ui.adsabs.harvard.edu/abs/2015ApJ...807L..30M}
}

@BOOK{Frank2002,
       author = {{Frank}, Juhan and {King}, Andrew and {Raine}, Derek J.},
        title = "{Accretion Power in Astrophysics: Third Edition}",
         year = 2002,
       adsurl = {https://ui.adsabs.harvard.edu/abs/2002apa..book.....F},
    publisher = {Cambridge University Press}
}

@ARTICLE{Ingram2011,
       author = {{Ingram}, Adam and {Done}, Chris},
        title = "{A physical model for the continuum variability and quasi-periodic oscillation in accreting black holes}",
      journal = {\mnras},
         year = 2011,
        month = aug,
       volume = {415},
       number = {3},
        pages = {2323-2335},
          doi = {10.1111/j.1365-2966.2011.18860.x},
archivePrefix = {arXiv},
       eprint = {1101.2336},
 primaryClass = {astro-ph.SR},
       adsurl = {https://ui.adsabs.harvard.edu/abs/2011MNRAS.415.2323I}
}

@ARTICLE{Nowak1999,
       author = {{Nowak}, Michael A. and {Vaughan}, Brian A. and {Wilms}, J{\"o}rn and {Dove}, James B. and {Begelman}, Mitchell C.},
        title = "{Rossi X-Ray Timing Explorer Observation of Cygnus X-1. II. Timing Analysis}",
      journal = {\apj},
         year = 1999,
        month = jan,
       volume = {510},
       number = {2},
        pages = {874-891},
          doi = {10.1086/306610},
archivePrefix = {arXiv},
       eprint = {astro-ph/9807278},
 primaryClass = {astro-ph},
       adsurl = {https://ui.adsabs.harvard.edu/abs/1999ApJ...510..874N}
}

@ARTICLE{Reig2003,
       author = {{Reig}, P. and {Kylafis}, N.~D. and {Giannios}, D.},
        title = "{Energy and time-lag spectra of galactic black-hole X-ray sources in the low/hard state}",
      journal = {\aap},
         year = 2003,
        month = may,
       volume = {403},
        pages = {L15-L18},
          doi = {10.1051/0004-6361:20030449},
archivePrefix = {arXiv},
       eprint = {astro-ph/0303585},
 primaryClass = {astro-ph},
       adsurl = {https://ui.adsabs.harvard.edu/abs/2003A&A...403L..15R}
}

@ARTICLE{Cassatella2012,
       author = {{Cassatella}, P. and {Uttley}, P. and {Wilms}, J. and {Poutanen}, J.},
        title = "{Joint spectral-timing modelling of the hard lags in GX 339-4: constraints on reflection models}",
      journal = {\mnras},
         year = 2012,
        month = may,
       volume = {422},
       number = {3},
        pages = {2407-2416},
          doi = {10.1111/j.1365-2966.2012.20792.x},
archivePrefix = {arXiv},
       eprint = {1202.4881},
 primaryClass = {astro-ph.HE},
       adsurl = {https://ui.adsabs.harvard.edu/abs/2012MNRAS.422.2407C}
}

@ARTICLE{Miyamoto1988,
       author = {{Miyamoto}, Sigenori and {Kitamoto}, Shunji and {Mitsuda}, Kazuhisa and {Dotani}, Tadayasu},
        title = "{Delayed hard X-rays from Cygnus X-l}",
      journal = {\nat},
         year = 1988,
        month = dec,
       volume = {336},
       number = {6198},
        pages = {450-452},
          doi = {10.1038/336450a0},
       adsurl = {https://ui.adsabs.harvard.edu/abs/1988Natur.336..450M}
}

@ARTICLE{Karpouzas2020,
       author = {{Karpouzas}, Konstantinos and {M{\'e}ndez}, Mariano and {Ribeiro}, Evandro M. and {Altamirano}, Diego and {Blaes}, Omer and {Garc{\'\i}a}, Federico},
        title = "{The Comptonizing medium of the neutron star in 4U 1636 - 53 through its lower kilohertz quasi-periodic oscillations}",
      journal = {\mnras},
         year = 2020,
        month = feb,
       volume = {492},
       number = {1},
        pages = {1399-1415},
          doi = {10.1093/mnras/stz3502},
archivePrefix = {arXiv},
       eprint = {1912.05380},
 primaryClass = {astro-ph.HE},
       adsurl = {https://ui.adsabs.harvard.edu/abs/2020MNRAS.492.1399K}
}

@ARTICLE{Karpouzas2021,
       author = {{Karpouzas}, Konstantinos and {M{\'e}ndez}, Mariano and {Garc{\'\i}a}, Federico and {Zhang}, Liang and {Altamirano}, Diego and {Belloni}, Tomaso and {Zhang}, Yuexin},
        title = "{A variable corona for GRS 1915+105}",
      journal = {\mnras},
         year = 2021,
        month = jun,
       volume = {503},
       number = {4},
        pages = {5522-5533},
          doi = {10.1093/mnras/stab827},
archivePrefix = {arXiv},
       eprint = {2103.09675},
 primaryClass = {astro-ph.HE},
       adsurl = {https://ui.adsabs.harvard.edu/abs/2021MNRAS.503.5522K}
}

@ARTICLE{Giannios2004,
       author = {{Giannios}, D. and {Kylafis}, N.~D. and {Psaltis}, D.},
        title = "{Spectra and time variability of Galactic black-hole X-ray sources in the low/hard state}",
      journal = {\aap},
         year = 2004,
        month = oct,
       volume = {425},
        pages = {163-169},
          doi = {10.1051/0004-6361:20041002},
archivePrefix = {arXiv},
       eprint = {astro-ph/0405569},
 primaryClass = {astro-ph},
       adsurl = {https://ui.adsabs.harvard.edu/abs/2004A&A...425..163G}
}

@ARTICLE{Giannios2005,
       author = {{Giannios}, D.},
        title = "{Spectra of black-hole binaries in the low/hard state: From radio to X-rays}",
      journal = {\aap},
         year = 2005,
        month = jul,
       volume = {437},
       number = {3},
        pages = {1007-1015},
          doi = {10.1051/0004-6361:20041491},
archivePrefix = {arXiv},
       eprint = {astro-ph/0504179},
 primaryClass = {astro-ph},
       adsurl = {https://ui.adsabs.harvard.edu/abs/2005A&A...437.1007G}
}

@BOOK{Kopal1959,
       author = {{Kopal}, Zdenek},
        title = "{Close binary systems}",
         year = 1959,
       adsurl = {https://ui.adsabs.harvard.edu/abs/1959cbs..book.....K}
}

@ARTICLE{Walker1956,
       author = {{Walker}, Merle F.},
        title = "{A Photometric Investigation of the Short-Period Eclipsing Binary, Nova DQ Herculis (1934).}",
      journal = {\apj},
         year = 1956,
        month = jan,
       volume = {123},
        pages = {68},
          doi = {10.1086/146132},
       adsurl = {https://ui.adsabs.harvard.edu/abs/1956ApJ...123...68W}
}

@ARTICLE{Walker1958,
       author = {{Walker}, Merle F.},
        title = "{Photoelectric Observations of Nova DQ Herculis (1934).}",
      journal = {\apj},
         year = 1958,
        month = mar,
       volume = {127},
        pages = {319},
          doi = {10.1086/146465},
       adsurl = {https://ui.adsabs.harvard.edu/abs/1958ApJ...127..319W}
}

@ARTICLE{Walker1961,
       author = {{Walker}, Merle F.},
        title = "{Photoelectric Observations of Nova (dq) Herculis, 1957-1959.}",
      journal = {\apj},
         year = 1961,
        month = jul,
       volume = {134},
        pages = {171},
          doi = {10.1086/147138},
       adsurl = {https://ui.adsabs.harvard.edu/abs/1961ApJ...134..171W}
}

@ARTICLE{Katz1975,
       author = {{Katz}, J.~I.},
        title = "{The structure of DQ Herculis.}",
      journal = {\apj},
         year = 1975,
        month = sep,
       volume = {200},
        pages = {298-305},
          doi = {10.1086/153788},
       adsurl = {https://ui.adsabs.harvard.edu/abs/1975ApJ...200..298K}
}

@ARTICLE{Chanmugam1977,
       author = {{Chanmugam}, G. and {Wagner}, R.~L.},
        title = "{The remarkable system AM Herculis / 3U 1809+50.}",
      journal = {\apjl},
         year = 1977,
        month = apr,
       volume = {213},
        pages = {L13-L16},
          doi = {10.1086/182399},
       adsurl = {https://ui.adsabs.harvard.edu/abs/1977ApJ...213L..13C}
}

@ARTICLE{King1979,
       author = {{King}, A.~R. and {Lasota}, J.~P.},
        title = "{Accretion on to highly magnetized white dwarfs.}",
      journal = {\mnras},
         year = 1979,
        month = sep,
       volume = {188},
        pages = {653-668},
          doi = {10.1093/mnras/188.3.653},
       adsurl = {https://ui.adsabs.harvard.edu/abs/1979MNRAS.188..653K}
}

@ARTICLE{Belloni1990,
       author = {{Belloni}, T. and {Hasinger}, G.},
        title = "{An atlas of aperiodic variability in HMXB.}",
      journal = {\aap},
         year = 1990,
        month = apr,
       volume = {230},
        pages = {103-119},
       adsurl = {https://ui.adsabs.harvard.edu/abs/1990A&A...230..103B}
}

@ARTICLE{PLATO,
       author = {{Rauer}, Heike and {Aerts}, Conny and {Cabrera}, Juan and {Deleuil}, Magali and {Erikson}, Anders and {Gizon}, Laurent and {Goupil}, Mariejo and {Heras}, Ana and {Walloschek}, Thomas and {Lorenzo-Alvarez}, Jose and {Marliani}, Filippo and {Martin-Garcia}, C{\'e}sar and {Mas-Hesse}, J. Miguel and {O'Rourke}, Laurence and {Osborn}, Hugh and {Pagano}, Isabella and {Piotto}, Giampaolo and {Pollacco}, Don and {Ragazzoni}, Roberto and {Ramsay}, Gavin and {Udry}, St{\'e}phane and {Appourchaux}, Thierry and {Benz}, Willy and {Brandeker}, Alexis and {G{\"u}del}, Manuel and {Janot-Pacheco}, Eduardo and {Kabath}, Petr and {Kjeldsen}, Hans and {Min}, Michiel and {Santos}, Nuno and {Smith}, Alan and {Suarez}, Juan-Carlos and {Werner}, Stephanie C. and {Aboudan}, Alessio and {Abreu}, Manuel and {Acu{\~n}a}, Lorena and {Adams}, Moritz and {Adibekyan}, Vardan and {Affer}, Laura and {Agneray}, Fran{\c{c}}ois and {Agnor}, Craig and {Aguirre B{\o}rsen-Koch}, Victor and {Ahmed}, Saad and {Aigrain}, Suzanne and {Al-Bahlawan}, Ashraf and {Alcacera Gil}, Ma de los Angeles and {Alei}, Eleonora and {Alencar}, Silvia and {Alexander}, Richard and {Alfonso-Garz{\'o}n}, Julia and {Alibert}, Yann and {Allende Prieto}, Carlos and {Almeida}, Leonardo and {Alonso Sobrino}, Roi and {Altavilla}, Giuseppe and {Althaus}, Christian and {Alvarez Trujillo}, Luis Alonso and {Amarsi}, Anish and {Ammler-von Eiff}, Matthias and {Am{\^o}res}, Eduardo and {Andrade}, Laerte and {Antoniadis-Karnavas}, Alexandros and {Ant{\'o}nio}, Carlos and {Aparicio del Moral}, Beatriz and {Appolloni}, Matteo and {Arena}, Claudio and {Armstrong}, David and {Aroca Aliaga}, Jose and {Asplund}, Martin and {Audenaert}, Jeroen and {Auricchio}, Natalia and {Avelino}, Pedro and {Baeke}, Ann and {Bailli{\'e}}, Kevin and {Balado}, Ana and {Ballber Balaguer{\'o}}, Pau and {Balestra}, Andrea and {Ball}, Warrick and {Ballans}, Herve and {Ballot}, Jerome and {Barban}, Caroline and {Barbary}, Ga{\"e}le and {Barbieri}, Mauro and {Barcel{\'o} Forteza}, Sebasti{\`a} and {Barker}, Adrian and {Barklem}, Paul and {Barnes}, Sydney and {Barrado Navascues}, David and {Barragan}, Oscar and {Baruteau}, Cl{\'e}ment and {Basu}, Sarbani and {Baudin}, Frederic and {Baumeister}, Philipp and {Bayliss}, Daniel and {Bazot}, Michael and {Beck}, Paul G. and {Belkacem}, Kevin and {Bellinger}, Earl and {Benatti}, Serena and {Benomar}, Othman and {B{\'e}rard}, Diane and {Bergemann}, Maria and {Bergomi}, Maria and {Bernardo}, Pierre and {Biazzo}, Katia and {Bignamini}, Andrea and {Bigot}, Lionel and {Billot}, Nicolas and {Binet}, Martin and {Biondi}, David and {Biondi}, Federico and {Birch}, Aaron C. and {Bitsch}, Bertram and {Bluhm Ceballos}, Paz Victoria and {B{\'o}di}, Attila and {Bogn{\'a}r}, Zs{\'o}fia and {Boisse}, Isabelle and {Bolmont}, Emeline and {Bonanno}, Alfio and {Bonavita}, Mariangela and {Bonfanti}, Andrea and {Bonfils}, Xavier and {Bonito}, Rosaria and {Bonomo}, Aldo Stefano and {B{\"o}rner}, Anko and {Boro Saikia}, Sudeshna and {Borreguero Mart{\'\i}n}, Elisa and {Borsa}, Francesco and {Borsato}, Luca and {Bossini}, Diego and {Bouchy}, Francois and {Bou{\'e}}, Gwena{\"e}l and {Boufleur}, Rodrigo and {Boumier}, Patrick and {Bourrier}, Vincent and {Bowman}, Dominic M. and {Bozzo}, Enrico and {Bradley}, Louisa and {Bray}, John and {Bressan}, Alessandro and {Breton}, Sylvain and {Brienza}, Daniele and {Brito}, Ana and {Brogi}, Matteo and {Brown}, Beverly and {Brown}, David J.~A. and {Brun}, Allan Sacha and {Bruno}, Giovanni and {Bruns}, Michael and {Buchhave}, Lars A. and {Bugnet}, Lisa and {Buldgen}, Ga{\"e}l and {Burgess}, Patrick and {Busatta}, Andrea and {Busso}, Giorgia and {Buzasi}, Derek and {Caballero}, Jos{\'e} A. and {Cabral}, Alexandre and {Cabrero Gomez}, Juan-Francisco and {Calderone}, Flavia and {Cameron}, Robert and {Cameron}, Andrew and {Campante}, Tiago and {Campos Gestal}, N{\'e}stor and {Canto Martins}, Bruno Leonardo and {Cara}, Christophe and {Carone}, Ludmila and {Carrasco}, Josep Manel and {Casagrande}, Luca and {Casewell}, Sarah L. and {Cassisi}, Santi and {Castellani}, Marco and {Castro}, Matthieu and {Catala}, Claude and {Catal{\'a}n Fern{\'a}ndez}, Irene and {Catelan}, M{\'a}rcio and {Cegla}, Heather and {Cerruti}, Chiara and {Cessa}, Virginie and {Chadid}, Merieme and {Chaplin}, William and {Charpinet}, Stephane and {Chiappini}, Cristina and {Chiarucci}, Simone and {Chiavassa}, Andrea and {Chinellato}, Simonetta and {Chirulli}, Giovanni and {Christensen-Dalsgaard}, J{\o}rgen and {Church}, Ross and {Claret}, Antonio and {Clarke}, Cathie and {Claudi}, Riccardo and {Clermont}, Lionel and {Coelho}, Hugo and {Coelho}, Joao and {Cogato}, Fabrizio and {Colom{\'e}}, Josep and {Condamin}, Mathieu and {Conde Garc{\'\i}a}, Fernando and {Conseil}, Simon},
        title = "{The PLATO mission}",
      journal = {Experimental Astronomy},
         year = 2025,
        month = jun,
       volume = {59},
       number = {3},
          eid = {26},
        pages = {26},
          doi = {10.1007/s10686-025-09985-9},
archivePrefix = {arXiv},
       eprint = {2406.05447},
 primaryClass = {astro-ph.IM},
       adsurl = {https://ui.adsabs.harvard.edu/abs/2025ExA....59...26R}
}

@software{LMFIT,
  author       = {Newville, Matthew and
                  Otten, Renee and
                  Nelson, Andrew and
                  Stensitzki, Till and
                  Ingargiola, Antonino and
                  Allan, Daniel and
                  Fox, Austin and
                  Carter, Faustin and
                  Rawlik, Michal},
  title        = {LMFIT: Non-Linear Least-Squares Minimization and
                   Curve-Fitting for Python
                  },
  month        = jul,
  year         = 2025,
  publisher    = {Zenodo},
  version      = {1.3.4},
  doi          = {10.5281/zenodo.16175987},
  url          = {https://doi.org/10.5281/zenodo.16175987},
  swhid        = {swh:1:dir:76742b0e41b1d2bff5a3716dd2376531f3a21db8
                   ;origin=https://doi.org/10.5281/zenodo.598352;visi
                   t=swh:1:snp:89f98ce93be53a573d85de0145f9174c827b7e
                   92;anchor=swh:1:rel:9528e5133d2d78c4036335f023b212
                   487cf1b632;path=lmfit-lmfit-py-0566445
                  },
}

@article{Bendat1986,
title = {Decomposition of wave forces into linear and non-linear components},
journal = {Journal of Sound and Vibration},
volume = {106},
number = {3},
pages = {391-408},
year = {1986},
issn = {0022-460X},
doi = {https://doi.org/10.1016/0022-460X(86)90186-0},
url = {https://www.sciencedirect.com/science/article/pii/0022460X86901860},
author = {J.S. Bendat and A.G. Piersol}
}

@ARTICLE{astropy,
       author = {{Astropy Collaboration} and {Price-Whelan}, Adrian M. and {Lim}, Pey Lian and {Earl}, Nicholas and {Starkman}, Nathaniel and {Bradley}, Larry and {Shupe}, David L. and {Patil}, Aarya A. and {Corrales}, Lia and {Brasseur}, C.~E. and {N{\"o}the}, Maximilian and {Donath}, Axel and {Tollerud}, Erik and {Morris}, Brett M. and {Ginsburg}, Adam and {Vaher}, Eero and {Weaver}, Benjamin A. and {Tocknell}, James and {Jamieson}, William and {van Kerkwijk}, Marten H. and {Robitaille}, Thomas P. and {Merry}, Bruce and {Bachetti}, Matteo and {G{\"u}nther}, H. Moritz and {Aldcroft}, Thomas L. and {Alvarado-Montes}, Jaime A. and {Archibald}, Anne M. and {B{\'o}di}, Attila and {Bapat}, Shreyas and {Barentsen}, Geert and {Baz{\'a}n}, Juanjo and {Biswas}, Manish and {Boquien}, M{\'e}d{\'e}ric and {Burke}, D.~J. and {Cara}, Daria and {Cara}, Mihai and {Conroy}, Kyle E. and {Conseil}, Simon and {Craig}, Matthew W. and {Cross}, Robert M. and {Cruz}, Kelle L. and {D'Eugenio}, Francesco and {Dencheva}, Nadia and {Devillepoix}, Hadrien A.~R. and {Dietrich}, J{\"o}rg P. and {Eigenbrot}, Arthur Davis and {Erben}, Thomas and {Ferreira}, Leonardo and {Foreman-Mackey}, Daniel and {Fox}, Ryan and {Freij}, Nabil and {Garg}, Suyog and {Geda}, Robel and {Glattly}, Lauren and {Gondhalekar}, Yash and {Gordon}, Karl D. and {Grant}, David and {Greenfield}, Perry and {Groener}, Austen M. and {Guest}, Steve and {Gurovich}, Sebastian and {Handberg}, Rasmus and {Hart}, Akeem and {Hatfield-Dodds}, Zac and {Homeier}, Derek and {Hosseinzadeh}, Griffin and {Jenness}, Tim and {Jones}, Craig K. and {Joseph}, Prajwel and {Kalmbach}, J. Bryce and {Karamehmetoglu}, Emir and {Ka{\l}uszy{\'n}ski}, Miko{\l}aj and {Kelley}, Michael S.~P. and {Kern}, Nicholas and {Kerzendorf}, Wolfgang E. and {Koch}, Eric W. and {Kulumani}, Shankar and {Lee}, Antony and {Ly}, Chun and {Ma}, Zhiyuan and {MacBride}, Conor and {Maljaars}, Jakob M. and {Muna}, Demitri and {Murphy}, N.~A. and {Norman}, Henrik and {O'Steen}, Richard and {Oman}, Kyle A. and {Pacifici}, Camilla and {Pascual}, Sergio and {Pascual-Granado}, J. and {Patil}, Rohit R. and {Perren}, Gabriel I. and {Pickering}, Timothy E. and {Rastogi}, Tanuj and {Roulston}, Benjamin R. and {Ryan}, Daniel F. and {Rykoff}, Eli S. and {Sabater}, Jose and {Sakurikar}, Parikshit and {Salgado}, Jes{\'u}s and {Sanghi}, Aniket and {Saunders}, Nicholas and {Savchenko}, Volodymyr and {Schwardt}, Ludwig and {Seifert-Eckert}, Michael and {Shih}, Albert Y. and {Jain}, Anany Shrey and {Shukla}, Gyanendra and {Sick}, Jonathan and {Simpson}, Chris and {Singanamalla}, Sudheesh and {Singer}, Leo P. and {Singhal}, Jaladh and {Sinha}, Manodeep and {Sip{\H{o}}cz}, Brigitta M. and {Spitler}, Lee R. and {Stansby}, David and {Streicher}, Ole and {{\v{S}}umak}, Jani and {Swinbank}, John D. and {Taranu}, Dan S. and {Tewary}, Nikita and {Tremblay}, Grant R. and {de Val-Borro}, Miguel and {Van Kooten}, Samuel J. and {Vasovi{\'c}}, Zlatan and {Verma}, Shresth and {de Miranda Cardoso}, Jos{\'e} Vin{\'\i}cius and {Williams}, Peter K.~G. and {Wilson}, Tom J. and {Winkel}, Benjamin and {Wood-Vasey}, W.~M. and {Xue}, Rui and {Yoachim}, Peter and {Zhang}, Chen and {Zonca}, Andrea and {Astropy Project Contributors}},
        title = "{The Astropy Project: Sustaining and Growing a Community-oriented Open-source Project and the Latest Major Release (v5.0) of the Core Package}",
      journal = {\apj},
         year = 2022,
        month = aug,
       volume = {935},
       number = {2},
          eid = {167},
        pages = {167},
          doi = {10.3847/1538-4357/ac7c74},
archivePrefix = {arXiv},
       eprint = {2206.14220},
 primaryClass = {astro-ph.IM},
       adsurl = {https://ui.adsabs.harvard.edu/abs/2022ApJ...935..167A}
}

@ARTICLE{SciPy2020,
       author = {{Virtanen}, Pauli and {Gommers}, Ralf and {Oliphant}, Travis E. and {Haberland}, Matt and {Reddy}, Tyler and {Cournapeau}, David and {Burovski}, Evgeni and {Peterson}, Pearu and {Weckesser}, Warren and {Bright}, Jonathan and {van der Walt}, St{\'e}fan J. and {Brett}, Matthew and {Wilson}, Joshua and {Millman}, K. Jarrod and {Mayorov}, Nikolay and {Nelson}, Andrew R.~J. and {Jones}, Eric and {Kern}, Robert and {Larson}, Eric and {Carey}, C.~J. and {Polat}, {\.I}lhan and {Feng}, Yu and {Moore}, Eric W. and {VanderPlas}, Jake and {Laxalde}, Denis and {Perktold}, Josef and {Cimrman}, Robert and {Henriksen}, Ian and {Quintero}, E.~A. and {Harris}, Charles R. and {Archibald}, Anne M. and {Ribeiro}, Ant{\^o}nio H. and {Pedregosa}, Fabian and {van Mulbregt}, Paul and {SciPy 1. 0 Contributors}},
        title = "{SciPy 1.0: fundamental algorithms for scientific computing in Python}",
      journal = {Nature Methods},
         year = 2020,
        month = feb,
       volume = {17},
        pages = {261-272},
          doi = {10.1038/s41592-019-0686-2},
archivePrefix = {arXiv},
       eprint = {1907.10121},
 primaryClass = {cs.MS},
       adsurl = {https://ui.adsabs.harvard.edu/abs/2020NatMe..17..261V}
}

@Article{Hunter2007,
  Author    = {Hunter, J. D.},
  Title     = {Matplotlib: A 2D graphics environment},
  Journal   = {Computing in Science \& Engineering},
  Volume    = {9},
  Number    = {3},
  Pages     = {90--95},
  publisher = {IEEE COMPUTER SOC},
  doi       = {10.1109/MCSE.2007.55},
  year      = 2007
}

@Article{         Harris2020,
 title         = {Array programming with {NumPy}},
 author        = {Charles R. Harris and K. Jarrod Millman and St{\'{e}}fan J.
                 van der Walt and Ralf Gommers and Pauli Virtanen and David
                 Cournapeau and Eric Wieser and Julian Taylor and Sebastian
                 Berg and Nathaniel J. Smith and Robert Kern and Matti Picus
                 and Stephan Hoyer and Marten H. van Kerkwijk and Matthew
                 Brett and Allan Haldane and Jaime Fern{\'{a}}ndez del
                 R{\'{i}}o and Mark Wiebe and Pearu Peterson and Pierre
                 G{\'{e}}rard-Marchant and Kevin Sheppard and Tyler Reddy and
                 Warren Weckesser and Hameer Abbasi and Christoph Gohlke and
                 Travis E. Oliphant},
 year          = {2020},
 month         = sep,
 journal       = {Nature},
 volume        = {585},
 number        = {7825},
 pages         = {357--362},
 doi           = {10.1038/s41586-020-2649-2},
 publisher     = {Springer Science and Business Media {LLC}},
 url           = {https://doi.org/10.1038/s41586-020-2649-2}
}

@ARTICLE{Lomb1976,
       author = {{Lomb}, N.~R.},
        title = "{Least-Squares Frequency Analysis of Unequally Spaced Data}",
      journal = {\apss},
         year = 1976,
        month = feb,
       volume = {39},
       number = {2},
        pages = {447-462},
          doi = {10.1007/BF00648343},
       adsurl = {https://ui.adsabs.harvard.edu/abs/1976Ap&SS..39..447L}
}

@ARTICLE{Scargle1982,
       author = {{Scargle}, J.~D.},
        title = "{Studies in astronomical time series analysis. II. Statistical aspects of spectral analysis of unevenly spaced data.}",
      journal = {\apj},
         year = 1982,
        month = dec,
       volume = {263},
        pages = {835-853},
          doi = {10.1086/160554},
       adsurl = {https://ui.adsabs.harvard.edu/abs/1982ApJ...263..835S}
}

@ARTICLE{Dobrotka2015,
       author = {{Dobrotka}, A. and {Mineshige}, S. and {Ness}, J.-U.},
        title = "{Rms-flux relation and fast optical variability simulations of the nova-like system MV Lyr}",
      journal = {\mnras},
         year = 2015,
        month = mar,
       volume = {447},
       number = {4},
        pages = {3162-3169},
          doi = {10.1093/mnras/stu2631},
archivePrefix = {arXiv},
       eprint = {1412.3719},
 primaryClass = {astro-ph.SR},
       adsurl = {https://ui.adsabs.harvard.edu/abs/2015MNRAS.447.3162D}
}

@ARTICLE{Scaringi2014,
       author = {{Scaringi}, Simone},
        title = "{A physical model for the flickering variability in cataclysmic variables}",
      journal = {\mnras},
         year = 2014,
        month = feb,
       volume = {438},
       number = {2},
        pages = {1233-1241},
          doi = {10.1093/mnras/stt2270},
archivePrefix = {arXiv},
       eprint = {1311.6814},
 primaryClass = {astro-ph.GA},
       adsurl = {https://ui.adsabs.harvard.edu/abs/2014MNRAS.438.1233S}
}

@ARTICLE{Balman2019,
       author = {{Balman}, {\c{S}}.},
        title = "{Disk structure of cataclysmic variables and broadband noise characteristics in comparison with XRBs}",
      journal = {Astronomische Nachrichten},
         year = 2019,
        month = may,
       volume = {340},
       number = {4},
        pages = {296-301},
          doi = {10.1002/asna.201913613},
       adsurl = {https://ui.adsabs.harvard.edu/abs/2019AN....340..296B}
}

@ARTICLE{Castro2024,
       author = {{Castro}, Angel and {Michel}, Ra{\'u}l and {Castro Segura}, Noel and {Altamirano}, Diego and {Tejada}, Carlos and {Herrera}, Joel and {Colorado}, Enrique and {Sierra}, Gerardo and {Altamirano-D{\'e}vora}, Liliana and {Echevarr{\'\i}a}, Juan and {Sloot}, Rasjied and {Wijnands}, Rudy and {Zavala}, Iv{\'a}n and {Rojas}, David and {Hern{\'a}ndez Santisteban}, Juan V. and {Vincentelli}, Federico and {Hern{\'a}ndez-Landa}, Javier A. and {Wang}, Song and {Fuentes}, Melissa and {Gandhi}, Poshak and {Silva-Cabrera}, Jos{\'e} S. and {Ram{\'\i}rez V{\'e}lez}, Julio and {Garc{\'\i}a}, Benjam{\'\i}n and {Guisa}, Gerardo and {G{\'o}mez Maqueo Chew}, Yilen and {Montalvo}, Felipe and {Valenzuela}, Francisco},
        title = "{First light simultaneous triple-channel optical observations of the OPTICAM system at the OAN-SPM}",
      journal = {\na},
         year = 2024,
        month = nov,
       volume = {112},
          eid = {102262},
        pages = {102262},
          doi = {10.1016/j.newast.2024.102262},
       adsurl = {https://ui.adsabs.harvard.edu/abs/2024NewA..11202262C}
}

@ARTICLE{Massey1988,
       author = {{Massey}, Philip and {Strobel}, Kevin and {Barnes}, Jeannette V. and {Anderson}, Edwin},
        title = "{Spectrophotometric Standards}",
      journal = {\apj},
         year = 1988,
        month = may,
       volume = {328},
        pages = {315},
          doi = {10.1086/166294},
       adsurl = {https://ui.adsabs.harvard.edu/abs/1988ApJ...328..315M}
}

@ARTICLE{Revnivtsev2010,
       author = {{Revnivtsev}, M. and {Burenin}, R. and {Bikmaev}, I. and {Kniazev}, A. and {Buckley}, D.~A.~H. and {Pretorius}, M.~L. and {Khamitov}, I. and {Ak}, T. and {Eker}, Z. and {Melnikov}, S. and {Crawford}, S. and {Pavlinsky}, M.},
        title = "{Aperiodic optical variability of intermediate polars - cataclysmic variables with truncated accretion disks}",
      journal = {\aap},
         year = 2010,
        month = apr,
       volume = {513},
          eid = {A63},
        pages = {A63},
          doi = {10.1051/0004-6361/200913355},
archivePrefix = {arXiv},
       eprint = {1002.4073},
 primaryClass = {astro-ph.GA},
       adsurl = {https://ui.adsabs.harvard.edu/abs/2010A&A...513A..63R}
}

@INPROCEEDINGS{Tody1986,
       author = {{Tody}, Doug},
        title = "{The IRAF Data Reduction and Analysis System}",
    booktitle = {Instrumentation in astronomy VI},
         year = 1986,
       editor = {{Crawford}, David L.},
       series = {Society of Photo-Optical Instrumentation Engineers (SPIE) Conference Series},
       volume = {627},
        month = jan,
        pages = {733},
          doi = {10.1117/12.968154},
       adsurl = {https://ui.adsabs.harvard.edu/abs/1986SPIE..627..733T}
}

@INPROCEEDINGS{Tody1993,
       author = {{Tody}, Doug},
        title = "{IRAF in the Nineties}",
    booktitle = {Astronomical Data Analysis Software and Systems II},
         year = 1993,
       editor = {{Hanisch}, R.~J. and {Brissenden}, R.~J.~V. and {Barnes}, J.},
       series = {Astronomical Society of the Pacific Conference Series},
       volume = {52},
        month = jan,
        pages = {173},
       adsurl = {https://ui.adsabs.harvard.edu/abs/1993ASPC...52..173T}
}

@ARTICLE{Saito2010,
       author = {{Saito}, R.~K. and {Baptista}, R. and {Horne}, K. and {Martell}, P.},
        title = "{Spectral Mapping of the Intermediate Polar DQ Herculis}",
      journal = {\aj},
         year = 2010,
        month = jun,
       volume = {139},
       number = {6},
        pages = {2542-2556},
          doi = {10.1088/0004-6256/139/6/2542},
archivePrefix = {arXiv},
       eprint = {1005.1612},
 primaryClass = {astro-ph.SR},
       adsurl = {https://ui.adsabs.harvard.edu/abs/2010AJ....139.2542S}
}

@ARTICLE{ODonnell1994,
       author = {{O'Donnell}, James E.},
        title = "{R v-dependent Optical and Near-Ultraviolet Extinction}",
      journal = {\apj},
         year = 1994,
        month = feb,
       volume = {422},
        pages = {158},
          doi = {10.1086/173713},
       adsurl = {https://ui.adsabs.harvard.edu/abs/1994ApJ...422..158O}
}

@ARTICLE{Bastiaansen1992,
       author = {{Bastiaansen}, P.~A.},
        title = "{Narrow band multicolor photometry of reddened and unreddened early-type stars.}",
      journal = {\aaps},
         year = 1992,
        month = jun,
       volume = {93},
        pages = {449-462},
       adsurl = {https://ui.adsabs.harvard.edu/abs/1992A&AS...93..449B}
}

@ARTICLE{Predehl1995,
       author = {{Predehl}, P. and {Schmitt}, J.~H.~M.~M.},
        title = "{X-raying the interstellar medium: ROSAT observations of dust scattering halos.}",
      journal = {\aap},
         year = 1995,
        month = jan,
       volume = {293},
        pages = {889-905},
       adsurl = {https://ui.adsabs.harvard.edu/abs/1995A&A...293..889P}
}

@software{Barbary2016,
  author       = {Barbary, Kyle},
  title        = {extinction v0.3.0},
  month        = dec,
  year         = 2016,
  publisher    = {Zenodo},
  doi          = {10.5281/zenodo.804967},
  url          = {https://doi.org/10.5281/zenodo.804967},
}

@ARTICLE{Scaringi2012,
       author = {{Scaringi}, S. and {K{\"o}rding}, E. and {Uttley}, P. and {Groot}, P.~J. and {Knigge}, C. and {Still}, M. and {Jonker}, P.},
        title = "{Broad-band timing properties of the accreting white dwarf MV Lyrae}",
      journal = {\mnras},
         year = 2012,
        month = dec,
       volume = {427},
       number = {4},
        pages = {3396-3405},
          doi = {10.1111/j.1365-2966.2012.22022.x},
archivePrefix = {arXiv},
       eprint = {1208.6292},
 primaryClass = {astro-ph.SR},
       adsurl = {https://ui.adsabs.harvard.edu/abs/2012MNRAS.427.3396S}
}

@ARTICLE{Shappee2014,
       author = {{Shappee}, B.~J. and {Prieto}, J.~L. and {Grupe}, D. and {Kochanek}, C.~S. and {Stanek}, K.~Z. and {De Rosa}, G. and {Mathur}, S. and {Zu}, Y. and {Peterson}, B.~M. and {Pogge}, R.~W. and {Komossa}, S. and {Im}, M. and {Jencson}, J. and {Holoien}, T.~W. -S. and {Basu}, U. and {Beacom}, J.~F. and {Szczygie{\l}}, D.~M. and {Brimacombe}, J. and {Adams}, S. and {Campillay}, A. and {Choi}, C. and {Contreras}, C. and {Dietrich}, M. and {Dubberley}, M. and {Elphick}, M. and {Foale}, S. and {Giustini}, M. and {Gonzalez}, C. and {Hawkins}, E. and {Howell}, D.~A. and {Hsiao}, E.~Y. and {Koss}, M. and {Leighly}, K.~M. and {Morrell}, N. and {Mudd}, D. and {Mullins}, D. and {Nugent}, J.~M. and {Parrent}, J. and {Phillips}, M.~M. and {Pojmanski}, G. and {Rosing}, W. and {Ross}, R. and {Sand}, D. and {Terndrup}, D.~M. and {Valenti}, S. and {Walker}, Z. and {Yoon}, Y.},
        title = "{The Man behind the Curtain: X-Rays Drive the UV through NIR Variability in the 2013 Active Galactic Nucleus Outburst in NGC 2617}",
      journal = {\apj},
         year = 2014,
        month = jun,
       volume = {788},
       number = {1},
          eid = {48},
        pages = {48},
          doi = {10.1088/0004-637X/788/1/48},
archivePrefix = {arXiv},
       eprint = {1310.2241},
 primaryClass = {astro-ph.HE},
       adsurl = {https://ui.adsabs.harvard.edu/abs/2014ApJ...788...48S}
}

@ARTICLE{Kochanek2017,
       author = {{Kochanek}, C.~S. and {Shappee}, B.~J. and {Stanek}, K.~Z. and {Holoien}, T.~W. -S. and {Thompson}, Todd A. and {Prieto}, J.~L. and {Dong}, Subo and {Shields}, J.~V. and {Will}, D. and {Britt}, C. and {Perzanowski}, D. and {Pojma{\'n}ski}, G.},
        title = "{The All-Sky Automated Survey for Supernovae (ASAS-SN) Light Curve Server v1.0}",
      journal = {\pasp},
         year = 2017,
        month = oct,
       volume = {129},
       number = {980},
        pages = {104502},
          doi = {10.1088/1538-3873/aa80d9},
archivePrefix = {arXiv},
       eprint = {1706.07060},
 primaryClass = {astro-ph.SR},
       adsurl = {https://ui.adsabs.harvard.edu/abs/2017PASP..129j4502K}
}




\appendix

\section{Additional fits}\label{appendix: extra fits}

\begin{figure*}
    \centering
    \includegraphics[width=\textwidth]{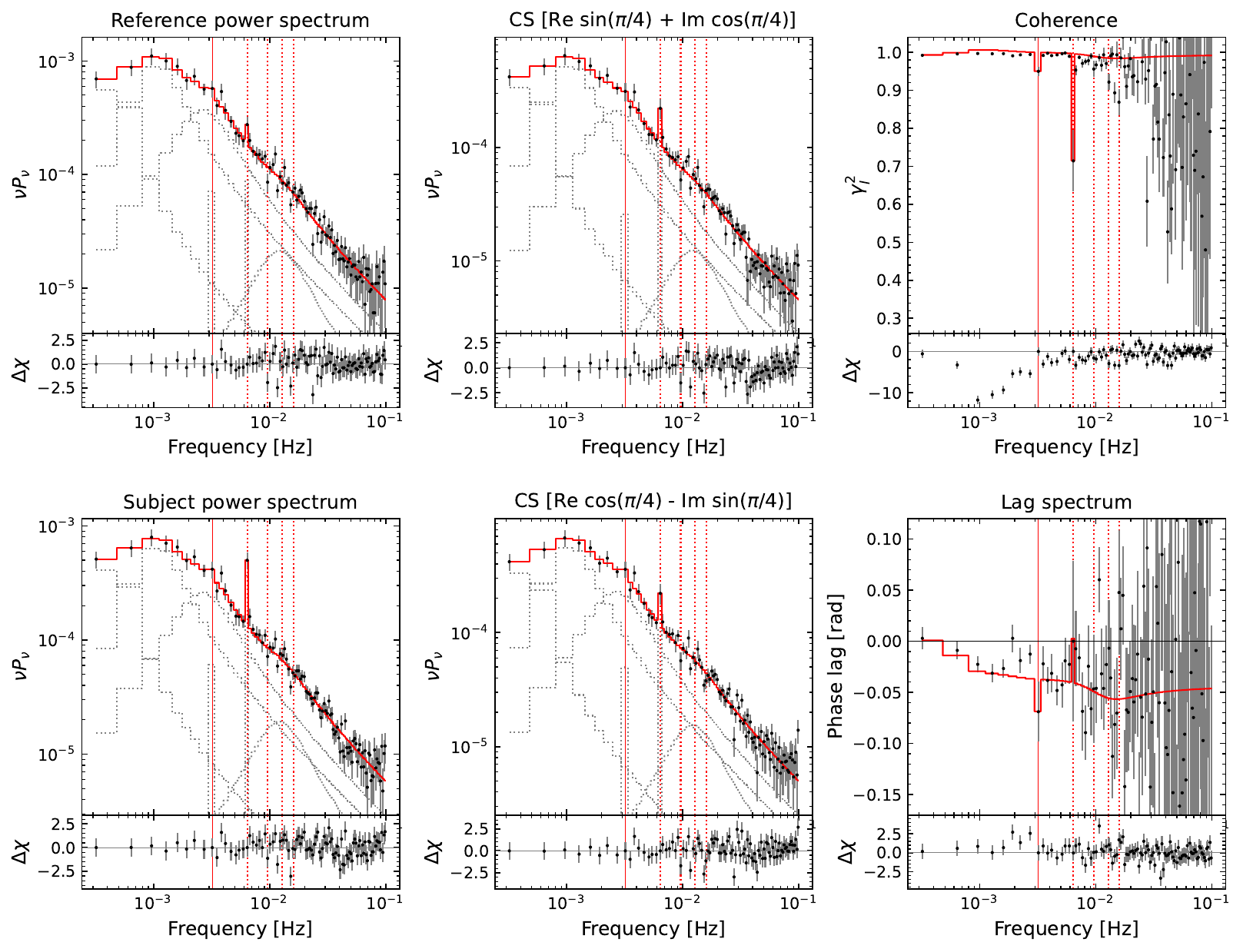}
    \caption{Similar to Figure \ref{fig: gxr 3 lor fit} but using four free Lorentzians instead of three.}
    \label{fig: gxr 4 lor fit}
\end{figure*}

\begin{figure*}
    \centering
    \includegraphics[width=\textwidth]{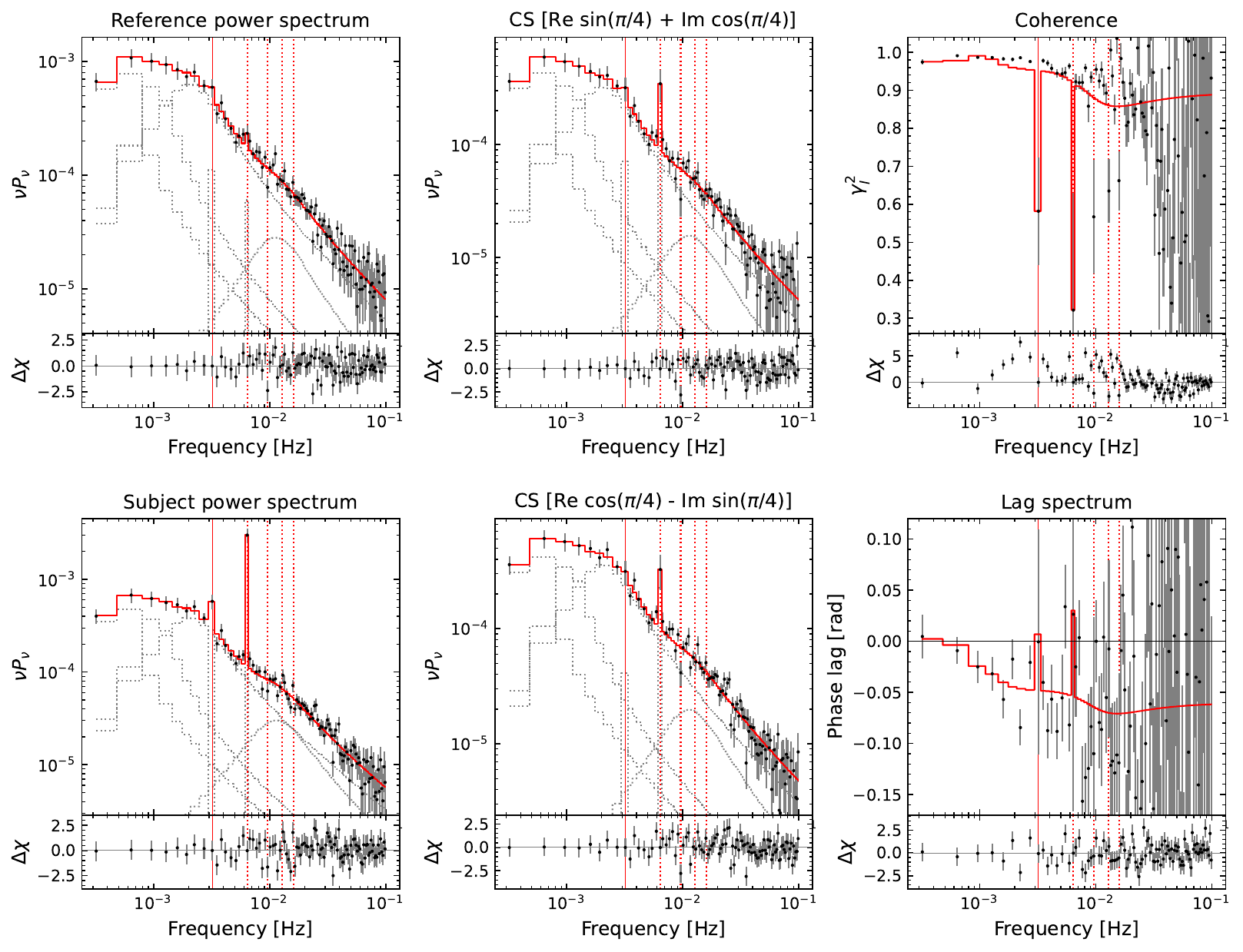}
    \caption{Similar to Figure \ref{fig: gxi 3 lor fit} but using four free Lorentzians instead of three.}
    \label{fig: gxi 4 lor fit}
\end{figure*}

\begin{figure*}
    \centering
    \includegraphics[width=\textwidth]{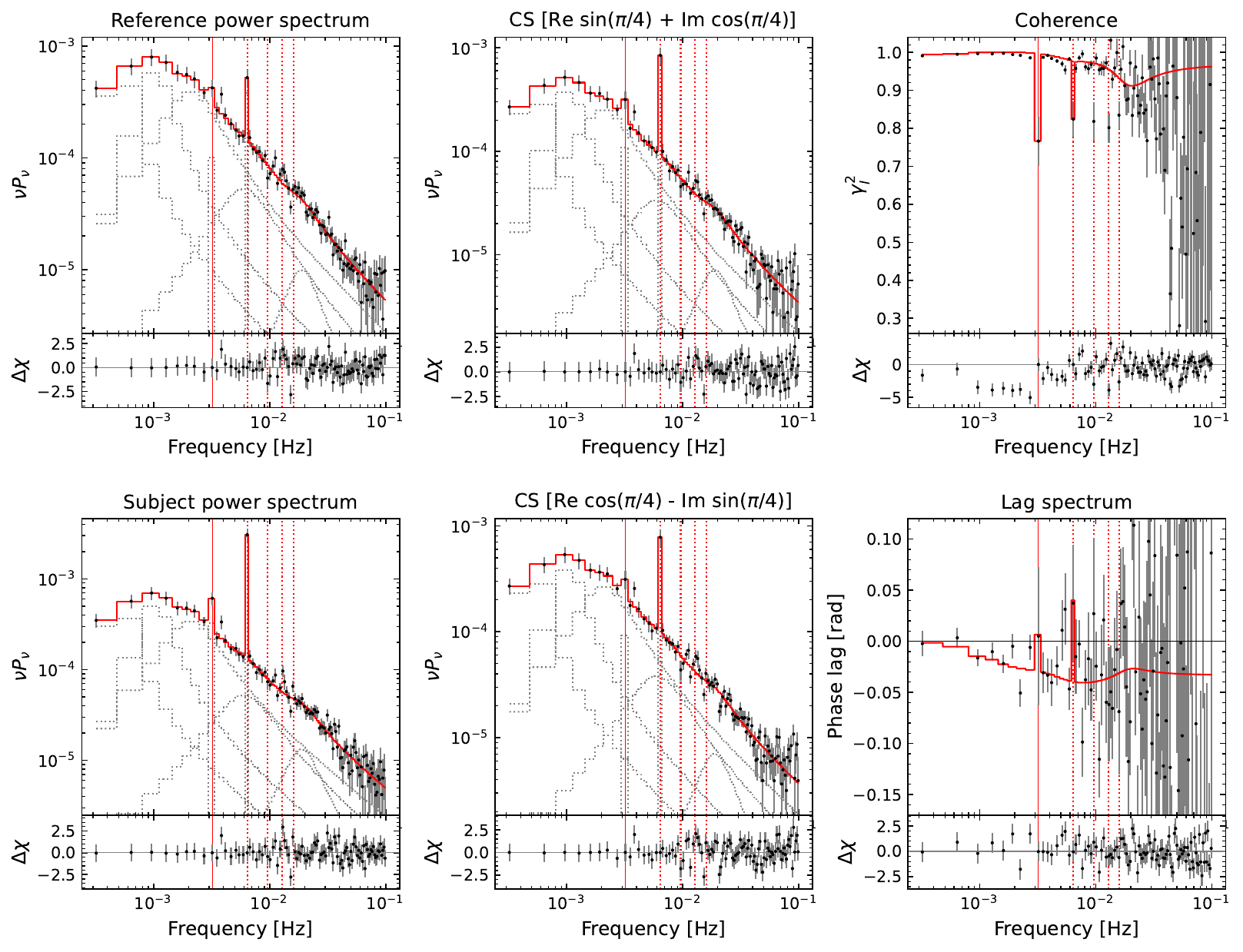}
    \caption{Similar to Figure \ref{fig: rxi 4 lor fit} but using five free Lorentzians instead of four.}
    \label{fig: rxi 5 lor fit}
\end{figure*}

In this section, we present additional fits to the cross and power spectra of our OPTICAM light curves. In Figure \ref{fig: gxr 4 lor fit}/Table \ref{tab: gxr 4 lor fit}, we present the four-Lorentzian fit to the cross and power spectra of the $g$- and $r$-bands; in Figure \ref{fig: gxi 4 lor fit}/Table \ref{tab: gxi 4 lor fit}, we present the four-Lorentzian fit to the cross and power spectra of the $g$- and $i$-bands; in Figure \ref{fig: rxi 5 lor fit}/Table \ref{tab: rxi 5 lor fit}, we present the five-Lorentzian fit to the cross and power spectra of the $r$- and $i$-bands. Each of these fits includes a Lorentzian that improves the recovery of the low-frequency coherence and lag spectra while not being significant in the cross or power spectra.

\begin{table*}
    \centering
    \caption{Four-Lorentzian fit to the cross and power spectra of the $g$- and $r$-bands shown in Figure \ref{fig: gxr 4 lor fit}.}
    \begin{tabular}{c|c|c|c|c|c|c|c}
        \hline
        Component & \multicolumn{3}{|c|}{rms [\%]} & $\nu_0$ [mHz] & $Q$ & Lag [rad] \\
         & PS$_\text{ref}$ & PS$_\text{sub}$ & CS \\
        \hline
        1 & $4^{*}_{- 2}$ & $4^{*}_{- 2}$ & $3^{*}_{- 1}$ & $0.4^{+ 0.6}_{- 0.1}$ & $5$ & $-0.8 \pm 0.1$ \\
        2 & $3.6^{+ 0.5}_{- 0.6}$ & $3.0^{+ 0.4}_{- 0.5}$ & $2.8^{+ 0.4}_{- 0.5}$ & $1.0^{*}_{- 0.1}$ & $0.9^{+ 0.4}_{- 0.2}$ & $-0.8 \pm 0.1$ \\
        3 & $2.5 \pm 0.5$ & $2.1 \pm 0.4$ & $1.9 \pm 0.4$ & $2.1^{+ 0.4}_{- 0.3}$ & $0.7^{+ 0.2}_{*}$ & $-0.83^{+ 0.08}_{- 0.07}$ \\
        4 & $0.6 \pm 0.2$ & $0.5 \pm 0.2$ & $0.5^{+ 0.2}_{- 0.1}$ & $10 \pm 3$ & $0.8^{+ 0.3}_{- 0.2}$ & $-0.9 \pm 0.1$ \\
        \hline
    \end{tabular}
    {
    \newline
    NOTES: $*$: unconstrained bound. All uncertainties are given to $1 \sigma$.
    }
    \label{tab: gxr 4 lor fit}
\end{table*}

\begin{table*}
    \centering
    \caption{Four-Lorentzian fit to the cross and power spectra of the $g$- and $i$-bands shown in Figure \ref{fig: gxi 4 lor fit}.}
    \begin{tabular}{c|c|c|c|c|c|c|c}
        \hline
        Component & \multicolumn{3}{|c|}{rms [\%]} & $\nu_0$ [mHz] & $Q$ & Lag [rad] \\
         & PS$_\text{ref}$ & PS$_\text{sub}$ & CS \\
        \hline
        1 & $0.7 \pm 0.2$ & $0.7^{+ 0.2}_{- 0.1}$ & $0.6^{+ 0.2}_{- 0.1}$ & $9 \pm 2$ & $0.6 \pm 0.2$ & $-0.9 \pm 0.1$ \\
        2 & $3.2^{+ 0.5}_{- 0.6}$ & $2.5^{+ 0.4}_{- 0.5}$ & $2.4^{+ 0.3}_{- 0.5}$ & $1.6^{+ 0.3}_{- 0.9}$ & $0.7^{+ 0.2}_{*}$ & $-0.84 \pm 0.06$ \\
        3 & $2.4^{+ 1.1}_{- 0.9}$ & $1.9^{+ 0.9}_{- 0.7}$ & $1.8^{+ 0.8}_{- 0.6}$ & $1.0^{+ 0.9}_{- 0.2}$ & $1.5^{*}_{- 0.8}$ & $-0.8 \pm 0.2$ \\
        4 & $7^{*}_{- 4}$ & $5^{*}_{- 3}$ & $5^{+ inf}_{- 3}$ & $0.46^{+ 0.01}_{- 0.03}$ & $9^{*}_{- 8}$ & $-0.8 \pm 0.1$ \\
        \hline
    \end{tabular}
    {
    \newline
    NOTES: $*$: unconstrained bound. All uncertainties are given to $1 \sigma$.
    }
    \label{tab: gxi 4 lor fit}
\end{table*}

\begin{table*}
    \centering
    \caption{Five-Lorentzian fit to the cross and power spectra of the $r$- and $i$-bands shown in Figure \ref{fig: rxi 5 lor fit}.}
    \begin{tabular}{c|c|c|c|c|c|c|c}
        \hline
        Component & \multicolumn{3}{|c|}{rms [\%]} & $\nu_0$ [mHz] & $Q$ & Lag [rad] \\
         & PS$_\text{ref}$ & PS$_\text{sub}$ & CS \\
        \hline
        1 & $5^{*}_{- 2}$ & $4^{*}_{- 2}$ & $4^{*}_{- 2}$ & $0.46^{+ 0.01}_{- 0.05}$ & $7^{*}_{- 6}$ & $-0.8 \pm 0.1$ \\
        2 & $2.3^{+ 0.9}_{- 0.7}$ & $2.1^{+ 0.8}_{- 0.6}$ & $1.9^{+ 0.7}_{- 0.5}$ & $1.01^{+ 0.09}_{- 0.18}$ & $1.7^{*}_{- 0.9}$ & $-0.8^{+ 0.2}_{- 0.1}$ \\
        3 & $2.5^{+ 0.5}_{*}$ & $2.3^{+ 0.4}_{- 0.6}$ & $2.0^{+ 0.4}_{*}$ & $1.7 \pm 0.3$ & $0.8^{+ 0.3}_{- 0.1}$ & $-0.81^{+ 0.09}_{- 0.08}$ \\
        4 & $0.9^{+ 0.4}_{- 0.3}$ & $0.9^{+ 0.4}_{- 0.3}$ & $0.8 \pm 0.3$ & $5^{+ 2}_{- 1}$ & $0.7^{+ 0.4}_{- 0.2}$ & $-0.8 \pm 0.1$ \\
        5 & $0.32^{+ 0.08}_{- 0.06}$ & $0.37^{+ 0.08}_{- 0.07}$ & $0.27^{+ 0.06}_{- 0.05}$ & $18^{+ 1}_{- 2}$ & $1.3^{+ 0.4}_{- 0.3}$ & $-0.8^{+ 0.2}_{- 0.1}$ \\
        \hline
    \end{tabular}
    {
    \newline
    NOTES: $*$: unconstrained bound. All uncertainties are given to $1 \sigma$.
    }
    \label{tab: rxi 5 lor fit}
\end{table*}

By comparing Figures \ref{fig: gxr 4 lor fit}, \ref{fig: gxi 4 lor fit} and \ref{fig: rxi 5 lor fit} to Figures \ref{fig: gxi 3 lor fit}. \ref{fig: gxr 3 lor fit} and \ref{fig: rxi 4 lor fit}, respectively, it can be seen that the intrinsic coherence and lag spectra are generally better reproduced with the inclusion of a non-significant low-frequency Lorentzian. The exception to this is the lag spectrum between the $g$- and $i$-bands, which could be reproduced using only significant Lorentzians (Section \ref{sec: gxi results}). However, we still find large ($\sim 10 \sigma$) residuals in the intrinsic coherence functions of Figures \ref{fig: gxr 4 lor fit}, \ref{fig: gxi 4 lor fit} and \ref{fig: rxi 5 lor fit}.

Large residuals for the intrinsic coherence function are likely a result of misestimation of the Poisson noise level (see Section \ref{sec: optical caveats}). Unlike the intrinsic coherence, however, the lag spectrum is independent of the Poisson noise level. We are therefore more accepting of large residuals in the intrinsic coherence functions, and place more emphasis on the lag spectrum recovery.


\bsp	
\label{lastpage}
\end{document}